\documentclass{article}   	
\usepackage{amsmath,amssymb,amsthm,bm,colortbl,float,graphicx,multirow}   
\usepackage[hyphens]{url}
\usepackage[colorlinks,citecolor=blue,urlcolor=blue]{hyperref}

\theoremstyle{plain}
\newtheorem{theorem}{Theorem}
\newtheorem{lemma}{Lemma} 

\theoremstyle{remark}
\newtheorem*{remark}{Remark} 

\pdfoutput=1
\allowdisplaybreaks
\def\T{\textsf{T}}

\newenvironment{newreferences}
               {\section*{References}
                \begin{list}{}{\setlength{\itemsep}{0pt}
                               \setlength{\parsep}{0pt}
                               \setlength{\labelwidth}{0pt}
                               \setlength{\leftmargin}{12pt}
                               \setlength{\labelsep}{0pt}}
                \setlength{\itemindent}{-12pt}
               }{\end{list}}

\title{A Game-Theoretic Analysis of \\ \textit{Baccara Chemin de Fer}, II}
\author{Stewart N. Ethier\thanks{Department of Mathematics, University of Utah, ethier@math.utah.edu}\ \ and Jiyeon Lee\thanks{Department of Statistics, Yeungnam University, leejy@yu.ac.kr}} 
\date{}

\begin{document}
\maketitle

\begin{abstract}
In a previous paper, we considered several models of the parlor game \textit{baccara chemin de fer}, including Model B2 (a $2\times2^{484}$ matrix game) and Model B3 (a $2^5\times2^{484}$ matrix game), both of which depend on a positive-integer parameter $d$, the number of decks.  The key to solving the game under Model B2 was what we called Foster's algorithm, which applies to additive $2\times2^n$ matrix games.  Here ``additive'' means that the payoffs are additive in the $n$ binary choices that comprise a player II pure strategy.

In the present paper, we consider analogous models of the casino game \textit{baccara chemin de fer} that take into account the $100\,\alpha$ percent commission on Banker (player II) wins, where $0\le\alpha\le1/10$.  Thus, the game now depends not just on the discrete parameter $d$ but also on a continuous parameter $\alpha$.  Moreover, the game is no longer zero sum.  To find all Nash equilibria under Model B2, we generalize Foster's algorithm to additive $2\times2^n$ bimatrix games.  We find that, with rare exceptions, the Nash equilibrium is unique.  We also obtain a Nash equilibrium under Model B3, based on Model B2 results, but here we are unable to prove uniqueness.
\medskip

\smallskip

\noindent \textbf{Classification}: MSC primary 91A05; secondary 91A60
\end{abstract}

\section{Introduction}

The parlor game \textit{baccara chemin de fer} was one of the motivating examples that led to the development of noncooperative two-person game theory (Borel, 1924).  We can classify game-theoretic models of \textit{baccara} in two ways.  First according to how the cards are dealt:
\begin{itemize}
\item Model A (with replacement).  Cards are dealt with replacement from a single deck.
\item Model B (without replacement).  Cards are dealt without replacement from a $d$-deck shoe.
\end{itemize}
And second according to the information available to Player and Banker about their own two-card hands:
\begin{itemize}
\item Model 1 (Player total, Banker total).  Each of Player and Banker sees the total of his own two-card hand but not its composition.
\item Model 2 (Player total, Banker composition).  Banker sees the composition of his own two-card hand while Player sees only his own total.
\item Model 3 (Player composition, Banker composition).  Each of Player and Banker sees the composition of his own two-card hand.
\end{itemize}
(We do not consider the fourth possibility.)  Under Model A1 \textit{baccara} is a $2\times 2^{88}$ matrix game, which was solved by Kemeny and Snell (1957).  Under Model B2 \textit{baccara} is a $2\times 2^{484}$ matrix game, which was solved in part by Downton and Lockwood (1975) and in full by Ethier and G\'amez (2013).  Under Model B3 \textit{baccara} is a $2^5\times2^{484}$ matrix game, which was solved in part by Ethier and G\'amez (2013).

Each of these works was concerned with the parlor game \textit{baccara chemin de fer}, in contrast to the casino game.  The rules of the parlor game, which also apply to the casino game, are as in Ethier and G\'amez (2013):  The role of Banker rotates among the players (counter-clockwise), changing hands after a Banker loss or when Banker chooses to relinquish his role.  Banker announces the amount he is willing to risk, and the total amount bet on Player's hand cannot exceed that amount.  After a Banker win, all winnings must be added to the bank unless Banker chooses to withdraw.  The game is played with six standard 52-card decks mixed together.  Denominations A, 2--9, 10, J, Q, K have values 1, 2--9, 0, 0, 0, 0, respectively, and suits are irrelevant.  The total of a hand, comprising two or three cards, is the sum of the values of the cards, modulo 10.  In other words, only the final digit of the sum is used to evaluate a hand.  Two cards are dealt face down to Player and two face down to Banker, and each looks only at his own hand.  The object of the game is to have the higher total (closer to 9) at the end of play.  A two-card total of 8 or 9 is a \textit{natural}.  If either hand is a natural, the game is over.  If neither hand is a natural, Player then has the option of drawing a third card.  If he exercises this option, his third card is dealt face up.  Next, Banker, observing Player's third card, if any, has the option of drawing a third card.  This completes the game, and the higher total wins.  Winning bets on Player's hand are paid by Banker at even odds.  Losing bets on Player's hand are collected by Banker.  Hands of equal total result in a tie or a \textit{push} (no money is exchanged).  Since several players can bet on Player's hand, Player's strategy is restricted.  He must draw on a two-card total of 4 or less and stand on a two-card total of 6 or 7.  When his two-card total is 5, he is free to stand or draw as he chooses.  (The decision is usually made by the player with the largest bet.)  Banker, on whose hand no one can bet, has no constraints on his strategy under classical rules.

There is one important additional rule in the casino game:  The house collects a five percent commission on Banker wins.  (This commission has been known to be as high as ten percent; see   Villiod (1906).)  Thus, our aim in the present paper is to generalize the aforementioned results to allow for a $100\,\alpha$ percent commission on Banker wins.  We will assume that $0\le\alpha\le1/10$.  This makes \textit{baccara chemin de fer} a bimatrix game instead of a matrix game, one that depends on a positive integer parameter $d$ (under Model B), the number of decks, as well as a continuous parameter $\alpha$ (under Model A or B), the commission on Banker wins.  In the case of Model A1 all Nash equilibria were identified in an unpublished paper by the authors (Ethier and Lee, 2013), assuming only $0\le\alpha<2/5$.  Under Model A1 and the present assumption ($0\le\alpha\le1/10$), the Nash equilibrium is unique for each $\alpha$.

There are also unimportant additional rules in the casino game.  Specifically, in modern casino \textit{baccara chemin de fer}, Banker's strategy is severely restricted.  With a few exceptions, these restrictions are benign, but because of the exceptions we ignore them entirely.

Ethier and G\'amez (2013) studied Models A2, A3, B1, B2, and B3 in the special case $\alpha=0$.  That was part I, and the present paper, with $0\le\alpha\le1/10$, is part II.

To keep the paper from becoming unduly long, we will focus our attention on Models B2 and B3, leaving the simpler models A2, A3, and B1 to the interested reader.  The key to solving the parlor game under Model B2 was what we called Foster's algorithm, which applies to additive $2\times2^n$ matrix games.  Foster (1964) called it a computer technique.  Here ``additive'' means that the payoffs are additive in the $n$ binary choices that comprise a player II pure strategy.  

In Section~\ref{sec:Foster} we generalize Foster's algorithm to additive $2\times2^n$ bimatrix games.  The generalization is not straightforward.  In Section~\ref{sec:B2} we show that, with rare exceptions, the Nash equilibrium is unique under Model B2.  Uniqueness is important because it ensures that optimal strategies are unambiguous.  The proof of uniqueness is computer assisted, with computations carried out in infinite precision using \textit{Mathematica}.  In Section~\ref{sec:B3} we obtain a Nash equilibrium under Model B3, based on Model B2 results, but here, just as for the parlor game, we are unable to prove uniqueness.

\section{Two Lemmas for Additive Bimatrix Games}\label{sec:Foster}

A reduction lemma for additive $m\times2^n$ matrix games was stated by Ethier and G\'amez (2013).  It had already been used implicitly by Kemeny and Snell (1957), Foster (1964), and Downton and Lockwood (1975).  Here we generalize to additive $m\times2^n$ bimatrix games.

\begin{lemma}[Reduction by strict dominance]\label{lem:reduction}
Let $m\ge2$ and $n\ge1$ and consider an $m\times 2^n$ bimatrix game of the following form.  Player I has $m$ pure strategies, labeled $0,1,\ldots,m-1$.  Player II has $2^n$ pure strategies, labeled by the subsets $T\subset\{1,2,\ldots,n\}$.   For $u=0,1,\ldots,m-1$, there exist probabilities $p_u(0)\ge0$, $p_u(1)>0$, \dots, $p_u(n)>0$ with $p_u(0)+p_u(1)+\cdots+p_u(n)=1$ together with a real number $b_u(0)$, and for $l=1,2,\ldots,n$, there exists a real $m\times2$ matrix
\begin{equation*}
\begin{pmatrix}
b_{0,0}(l)&b_{0,1}(l)\\
b_{1,0}(l)&b_{1,1}(l)\\
\vdots & \vdots \\
b_{m-1,0}(l)&b_{m-1,1}(l)\\
\end{pmatrix}.
\end{equation*}
Assume that the $m\times 2^n$ bimatrix game has player II payoff matrix $\bm B$ with $(u,T)$ entry given by
\begin{equation*}
b_{u,T}:=p_u(0)b_u(0)+\sum_{l\in T^c}p_u(l)b_{u,0}(l)+\sum_{l\in T}p_u(l)b_{u,1}(l)
\end{equation*}
for $u\in\{0,1,\ldots,m-1\}$ and $T\subset\{1,2,\ldots,n\}$.  Here $T^c:=\{1,2,\ldots,n\}-T$.

We define
\begin{align*}
T_0&:=\{l\in\{1,2,\ldots,n\}: b_{u,0}(l)>b_{u,1}(l){\rm\ for\ }u=0,1,\ldots,m-1\},\\
T_1&:=\{l\in\{1,2,\ldots,n\}: b_{u,0}(l)<b_{u,1}(l){\rm\ for\ }u=0,1,\ldots,m-1\},\\
T_*&:=\{1,2,\ldots,n\}-T_0-T_1,
\end{align*}
and put $n_*:=|T_*|$.

Then, given $T\subset\{1,2,\ldots,n\}$, player II's pure strategy $T$ is strictly dominated unless $T_1\subset T\subset T_1\cup T_*$.  Therefore, the $m\times 2^n$ bimatrix game can be reduced to an $m\times 2^{n_*}$ bimatrix game with no loss of Nash equilibria.
\end{lemma}

\begin{remark}
The game can be thought of as follows.  Player I chooses a pure strategy $u\in\{0,1,\ldots,m-1\}$.  Then Nature chooses a random variable $Z_u$ with distribution $P(Z_u=l)=p_u(l)$ for $l=0,1,\ldots,n$.  Given that $Z_u=0$, the game is over and player II's conditional expected payoff is $b_u(0)$.  If $Z_u\in\{1,2,\ldots,n\}$, then player II observes $Z_u$ (but not $u$) and based on this information chooses a ``move'' $j\in\{0,1\}$.  Given that $Z_u=l$ and player II chooses move $0$ (resp., move $1$), player II's conditional expected payoff is $b_{u,0}(l)$ (resp., $b_{u,1}(l)$).  Thus, player II's pure strategies can be identified with subsets $T\subset\{1,2,\ldots,n\}$, with player II choosing move $0$ if $Z_u\in T^c$ and move $1$ if $Z_u\in T$.  The lemma implies that, regardless of player I's strategy choice, it is optimal for player II to choose move $0$ if $Z_u\in T_0$ and move $1$ if $Z_u\in T_1$. 
\end{remark}

\begin{proof}
Suppose that the condition $T_1\subset T\subset T_1\cup T_*$ fails.  There are
two cases.  In case 1, there exists $l_0\in T_1$ with
$l_0\notin T$.  Here define $T':=T\cup\{l_0\}$.  In case 2, there exists
$l_0\in T$ with $l_0\notin T_1\cup T_*$ (so $l_0\in T_0$).  Here define
$T':=T-\{l_0\}$.  Then, for $u=0,1,\ldots,m-1$,
\begin{align*}\label{gamesEq1.17}
b_{u,T'}&=p_u(0)b_u(0)+\sum_{l\in (T')^c}p_u(l)b_{u,0}(l)+\sum_{l\in T'}p_u(l)b_{u,1}(l)\\
&=p_u(0)b_u(0)+\sum_{l\in T^c}p_u(l)b_{u,0}(l)+\sum_{l\in T}p_u(l)b_{u,1}(l)\\
&\qquad\qquad\qquad\!{}\pm p_u(l_0)(b_{u,1}(l_0)-b_{u,0}(l_0))\\ \noalign{\newpage}
&>p_u(0)b_u(0)+\sum_{l\in T^c}p_u(l)b_{u,0}(l)+\sum_{l\in T}p_u(l)b_{u,1}(l)\\
&=b_{u,T},
\end{align*}
where the $\pm$ sign is a plus sign in case 1 and a minus sign in case 2.  This tells us that player II's pure strategy $T$ is strictly dominated by pure strategy $T'$, as required.
\end{proof}

Ethier and G\'amez (2013) formulated Foster's (1964) algorithm for solving additive $2\times 2^n$ matrix games.  Here we generalize that result to additive $2\times 2^n$ bimatrix games.

\begin{lemma}[Foster's algorithm]\label{lem:Foster}
Let $n\ge1$ and consider a $2\times 2^n$ bimatrix game
of the following form.  Player I has two pure strategies, labeled $0$ and $1$.
Player II has $2^n$ pure strategies, labeled by the subsets
$T\subset\{1,2,\ldots,n\}$.  For $u=0,1$, there exist probabilities $p_u(0)\ge0$, $p_u(1)>0$, \dots, $p_u(n)>0$ with $p_u(0)+p_u(1)+\cdots+p_u(n)=1$ together with a real number
$b_u(0)$, and for $l=1,2,\ldots,n$, there exists a real $2\times2$ matrix
\begin{equation*}
\begin{pmatrix}
b_{0,0}(l)&b_{0,1}(l)\\
b_{1,0}(l)&b_{1,1}(l)\\
\end{pmatrix}.
\end{equation*}
Assume that the $2\times 2^n$ bimatrix game has payoff bimatrix $(\bm A,\bm B)$ with $(u,T)$ entry given by $(a_{u,T},b_{u,T})$, where $a_{u,T}$ is an arbitrary real number and 
\begin{equation*}
b_{u,T}:=p_u(0)b_u(0)+\sum_{l\in T^c}p_u(l)b_{u,0}(l)+\sum_{l\in T}p_u(l)b_{u,1}(l)
\end{equation*}
for $u\in\{0,1\}$ and $T\subset\{1,2,\ldots,n\}$. Here $T^c:=\{1,2,\ldots,n\}-T$.

We define
\begin{align*}
T_{0,0}&:=\{l\in\{1,2,\ldots,n\}: b_{0,0}(l)>b_{0,1}(l)\text{ and }b_{1,0}(l)>b_{1,1}(l)\},\nonumber\\
T_{0,1}&:=\{l\in\{1,2,\ldots,n\}: b_{0,0}(l)\ge b_{0,1}(l)\text{ and }b_{1,0}(l)\le b_{1,1}(l)\nonumber\\
&\qquad\qquad\qquad\qquad\quad\;\,\text{with at least one of these two inequalities strict}\},\nonumber\\
T_{1,0}&:=\{l\in\{1,2,\ldots,n\}: b_{0,0}(l)\le b_{0,1}(l)\text{ and }b_{1,0}(l)\ge b_{1,1}(l)\nonumber\\
&\qquad\qquad\qquad\qquad\quad\;\,\text{with at least one of these two inequalities strict}\},\nonumber\\
T_{1,1}&:=\{l\in\{1,2,\ldots,n\}: b_{0,0}(l)<b_{0,1}(l)\text{ and }b_{1,0}(l)<b_{1,1}(l)\},\nonumber
\end{align*}
and assume that $T_{0,0}\cup T_{0,1}\cup T_{1,0}\cup T_{1,1}=\{1,2,\ldots,n\}$.

$({\rm a})$ If player I uses the mixed strategy $(1-p,p)$ for some $p\in[0,1]$, then player II's unique best response is the pure strategy
\begin{equation}\label{T(p)}
T(p):=T_{1,1}\cup\{l\in T_{0,1}:p(l)<p\}\cup\{l\in T_{1,0}:p(l)>p\},
\end{equation}
where
\begin{equation*}
p(l):=\frac{p_0(l)[b_{0,1}(l)-b_{0,0}(l)]}{p_0(l)[b_{0,1}(l)-b_{0,0}(l)]-p_1(l)[b_{1,1}(l)-b_{1,0}(l)]},
\end{equation*}
provided $p$ does not belong to $\{p(l): l\in T_{0,1}\cup T_{1,0}\}$.  If $p=p(l)$ for exactly one choice of $l\in T_{0,1}\cup T_{1,0}$, namely $l'$, then player II's set of best responses is the set of mixtures of the two pure strategies $T(p)$, as in Equation~\eqref{T(p)}, and $T(p)\cup\{l'\}$.  If $p=p(l)$ for exactly two choices of $l\in T_{0,1}\cup T_{1,0}$, namely $l'$ and $l''$, then player II's set of best responses is the set of mixtures of the four pure strategies $T(p)$, as in Equation~\eqref{T(p)}, $T(p)\cup\{l'\}$, $T(p)\cup\{l''\}$, and $T(p)\cup\{l',l''\}$.

$({\rm b})$  For each $p\in[0,1]$ with $p\notin\{p(l): l\in T_{0,1}\cup T_{1,0}\}$, assume that $a_{0,T(p)}\ne a_{1,T(p)}$.  Assume also that $a_{0,T(0)}<a_{1,T(0)}$ and $a_{0,T(1)}>a_{1,T(1)}$.  Then every Nash equilibrium $(\bm p,\bm q)$ must have $\bm p=(1-p(l),p(l))$ for some $l\in T_{0,1}\cup T_{1,0}$.

$({\rm c})$ Under the assumptions of part $({\rm b})$, if $p=p(l)$ for exactly one choice of $l\in T_{0,1}\cup T_{1,0}$, namely $l'$, then every Nash equilibrium $(\bm p,\bm q)$ must have $\bm p=(1-p,p)$ and $\bm q$ with entries $1-q$ and $q\in[0,1]$ at the coordinates corresponding to player II pure strategies $T(p)$ and $T(p)\cup\{l'\}$ $(0$s elsewhere$)$, where
\begin{equation}\label{equalizing}
(1-q)a_{0,T(p)}+q\,a_{0,T(p)\cup\{l'\}}=(1-q)a_{1,T(p)}+q\,a_{1,T(p)\cup\{l'\}}.   
\end{equation}
$\bm q$ is called an equalizing strategy.

$({\rm d})$ Under the assumptions of part $({\rm b})$, if $p=p(l)$ for exactly two choices of $l\in T_{0,1}\cup T_{1,0}$, namely $l'$ and $l''$, then every Nash equilibrium $(\bm p,\bm q)$ must have $\bm p=(1-p,p)$ and $\bm q$ with entries $q,q',q'',q'''\in[0,1]$ $($with $q+q'+q''+q'''=1)$ at the coordinates corresponding to player II pure strategies $T(p)$, $T(p)\cup\{l'\}$, $T(p)\cup\{l''\}$, and $T(p)\cup\{l',l''\}$ $(0$s elsewhere$)$, where 
\begin{align*}
&q\, a_{0,T(p)}+q' a_{0,T(p)\cup\{l'\}}+q'' a_{0,T(p)\cup\{l''\}}+q''' a_{0,T(p)\cup\{l',l''\}} \\
&\quad{}=q\,a_{1,T(p)}+q'a_{1,T(p)\cup\{l'\}}+q'' a_{1,T(p)\cup\{l''\}}+q''' a_{1,T(p)\cup\{l',l''\}}.  
\end{align*}
Again, $\bm q$ is called an equalizing strategy.
\end{lemma}

\begin{remark}
(a) Lemma~\ref{lem:reduction} implies that every player II pure strategy $T$ that does not satisfy $T_{1,1}\subset T\subset T_{1,1}\cup T_{0,1}\cup T_{1,0}$ is strictly dominated.  Thus, the $2\times2^n$ bimatrix game can be reduced to a $2\times2^{n_*}$ bimatrix game, where $n_*:=|T_{0,1}\cup T_{1,0}|$, with no loss of Nash equilibria.

(b) The reason for referring to this lemma as an algorithm is that it gives straightforward conditions for determining all Nash equilibria.  These conditions primarily involve checking for equalizing strategies in a limited number of cases.
\end{remark}

\begin{proof}
$(\text{a})$  For $T(p)$ to be player II's unique best response, it must be the case that $T\mapsto(1-p)b_{0,T}+p\,b_{1,T}$ is uniquely maximized at $T=T(p)$.  Now, for arbitrary $T$ that excludes $l'$, the additivity of player II's payoffs implies that 
\begin{equation}\label{ineq1}
(1-p)b_{0,T\cup\{l'\}}+p\,b_{1,T\cup\{l'\}}>(1-p)b_{0,T}+p\,b_{1,T}
\end{equation}
if and only if 
\begin{equation}\label{ineq2}
(1-p)p_0(l')b_{0,1}(l')+p\,p_1(l')b_{1,1}(l')>(1-p)p_0(l')b_{0,0}(l')+p\,p_1(l')b_{1,0}(l').
\end{equation} 
But Inequality~\eqref{ineq2} is equivalent to
\begin{equation}\label{ineq3}
(1-p)p_0(l')[b_{0,1}(l')-b_{0,0}(l')]+p\,p_1(l')[b_{1,1}(l')-b_{1,0}(l')]>0,
\end{equation}  
which holds if and only if
\begin{equation*}
l'\in T_{1,1}\cup\{l\in T_{0,1}: p(l)<p\}\cup\{l\in T_{1,0}: p(l)>p\}=:T(p).  
\end{equation*}
Similarly, 
\begin{equation*}
(1-p)b_{0,T\cup\{l'\}}+p\,b_{1,T\cup\{l'\}}<(1-p)b_{0,T}+p\,b_{1,T}
\end{equation*}
if and only if
\begin{equation}\label{ineq4}
l'\in T_{0,0}\cup\{l\in T_{0,1}: p(l)>p\}\cup\{l\in T_{1,0}: p(l)<p\}.  
\end{equation}
If we assume that $p\notin \{p(l): l\in T_{0,1}\cup T_{1,0}\}$, then Inclusion~\eqref{ineq4} is equivalent to $l'\in T(p)^c$.
The first conclusion of part $(\text{a})$ follows.  For the second conclusion, notice that Inequalities~\eqref{ineq1}--\eqref{ineq3}, with the inequalities replaced by equalities, are equivalent to each other and to $p(l')=p$.  This suffices.  The third conclusion follows similarly.

$(\text{b})$  For $p\in[0,1]$ with $p\notin\{p(l): l\in T_{0,1}\cup T_{1,0}\}$, we have seen that the pure strategy $T(p)$ is the unique best response to $\bm p=(1-p,p)$. However, for $0<p<1$, the mixed strategy $\bm p$ cannot be a best response to the pure strategy $T(p)$ unless $a_{0,T(p)}=a_{1,T(p)}$, which has been ruled out.  To extend this to $p=0$ and $p=1$, we note that neither $(0,T(0))$ nor $(1,T(1))$ is a pure Nash equilibrium, by virtue of the other assumptions of part $(\text{b})$.

$(\text{c})$  We assume that $p=p(l')$ for a unique $l'\in T_{0,1}\cup T_{1,0}$.  By part (a), any mixture of the pure strategies $T(p)$ and $T(p)\cup\{l'\}$ will be a best response to the mixed strategy $\bm p=(1-p,p)$, but at most one such mixture, namely the equalizing strategy that chooses $T(p)$ with probability $1-q$ and $T(p)\cup\{l'\}$ with probability $q$, where $q$ satisfies Equation~\eqref{equalizing}, will result in a Nash equilibrium.

$({\rm d})$ The proof is similar to that of part $({\rm c})$.
\end{proof}

\section{Model B2}\label{sec:B2}

In this section we study Model B2.  Here cards are dealt without replacement from a $d$-deck shoe, and Player sees only the total of his two-card hand, while Banker sees the composition of his two-card hand.  Player has a stand-or-draw decision at two-card totals of 5, and Banker has a stand-or-draw decision in $44\times11 = 484$ situations (44 compositions corresponding to Banker totals of 0--7, and 11 Player third-card values, 0--9 and $\varnothing$), so \textit{baccara chemin de fer} is a $2\times2^{484}$ bimatrix game.

We denote Player's two-card hand by $(X_1,X_2)$, where $0\le X_1\le X_2\le9$, Banker's two-card hand by $(Y_1,Y_2)$, where $0\le Y_1\le Y_2\le9$, and Player's and Banker's third-card values (possibly $\varnothing$) by $X_3$ and $Y_3$.  Note, for example, that $X_1$ and $X_2$ are not Player's first- and second-card values; rather, they are the smaller and larger values of Player's first two cards.  We define the function $M$ on the set of nonnegative integers by 
\begin{equation}\label{M}
M(i):=\text{mod}(i,10),
\end{equation} 
that is, $M(i)$ is the remainder when $i$ is divided by 10.  We define Player's two-card total by $X:=M(X_1+X_2)$.  We further denote by $G_0$ and $G_1$ Banker's profit in the parlor game from standing and from drawing, respectively, assuming a one-unit bet.  More precisely, for $v=0$ (Banker stands) and $v=1$ (Banker draws),
\begin{equation*}
G_v:=\begin{cases}1&\text{if Banker wins},\\  0&\text{if a tie occurs}, \\ -1&\text{if Player wins}.\end{cases}
\end{equation*}

We next define the relevant probabilities when cards are dealt without replacement.  Let
\begin{align}\label{p4}
p_4((i_1,i_2),(j_1,j_2))&:=(2-\delta_{i_1,i_2})\frac{4d(1+3\delta_{i_1,0})}{52d}\cdot\frac{4d(1+3\delta_{i_2,0})-\delta_{i_2,i_1}}{52d-1}\nonumber\\
&\qquad{}\cdot(2-\delta_{j_1,j_2})\frac{4d(1+3\delta_{j_1,0})-\delta_{j_1,i_1}-\delta_{j_1,i_2}}{52d-2}\nonumber\\
&\qquad{}\cdot\frac{4d(1+3\delta_{j_2,0})-\delta_{j_2,i_1}-\delta_{j_2,i_2}-\delta_{j_2,j_1}}{52d-3}
\end{align}
be the probability that Player's two-card hand is $(i_1,i_2)$, where $0\le i_1\le i_2\le9$, and Banker's two-card hand is $(j_1,j_2)$, where $0\le j_1\le j_2\le9$.  To elaborate on this formula, we note that the order of the cards within a two-card hand is irrelevant, so the hand comprising $i_1$ and $i_2$ can be written as $(\min(i_1,i_2),\max(i_1,i_2))$, and the factor $(2-\delta_{i_1,i_2})$ adjusts the probability accordingly.  The factors of the form $(1+3\delta_{i_1,0})$ take into account the fact that cards valued as 0 are four times as frequent as cards valued as 1, for example.  Finally, the terms of the form ${}-\delta_{i_2,i_1}$ are the effects of previously dealt cards. In practice, the order in which the first four cards are dealt is Player, Banker, Player, Banker.  But it is mathematically equivalent, and slightly more convenient, to assume that the order is Player, Player, Banker, Banker.  

Second,
\begin{align}\label{p5}
p_5((i_1,i_2),(j_1,j_2),k)&:=p_4((i_1,i_2),(j_1,j_2))\nonumber\\
&\qquad{}\cdot\frac{4d(1+3\delta_{k,0})-\delta_{k,i_1}-\delta_{k,i_2}-\delta_{k,j_1}-\delta_{k,j_2}}{52d-4}
\end{align}
is the probability that Player's two-card hand is $(i_1,i_2)$, where $0\le i_1\le i_2\le9$ and $M(i_1+i_2)\le7$,  Banker's two-card hand is $(j_1,j_2)$, where $0\le j_1\le j_2\le 9$ and $M(j_1+j_2)\le7$, and the fifth card dealt has value $k\in\{0,1,\ldots,9\}$.  Note that $\sum_{0\le k\le9}p_5((i_1,i_2),(j_1,j_2),k)=p_4((i_1,i_2),(j_1,j_2))$.

Third, 
\begin{align}\label{p6}
p_6((i_1,i_2),(j_1,j_2),k,l)&:=p_5((i_1,i_2),(j_1,j_2),k)\nonumber\\
&\qquad{}\cdot\frac{4d(1+3\delta_{l,0})-\delta_{l,i_1}-\delta_{l,i_2}-\delta_{l,j_1}-\delta_{l,j_2}-\delta_{l,k}}{52d-5}
\end{align}
is the probability that Player's two-card hand is $(i_1,i_2)$, where $0\le i_1\le i_2\le9$ and $M(i_1+i_2)\le7$,  Banker's two-card hand is $(j_1,j_2)$, where $0\le j_1\le j_2\le 9$ and $M(j_1+j_2)\le7$, the fifth card dealt has value $k\in\{0,1,\ldots,9\}$, and the sixth card dealt has value $l\in\{0,1,\ldots,9\}$.  Note that $\sum_{0\le l\le9}p_6((i_1,i_2),(j_1,j_2),k,l)=p_5((i_1,i_2),(j_1,j_2),k)$.

Given a function $f$ on the set of integers, let us define, for $u\in\{0,1\}$, $0\le j_1\le j_2\le9$ with $M(j_1+j_2)\le7$, and $k\in\{0,1,\ldots,9\}$,
\begin{align}\label{eu0(j1,j2,k)-B2}
e_{u,0}((j_1,j_2),k)&:=\sum_{i=0}^{4+u}\;\sum_{\substack{0\le i_1\le i_2\le 9:\\ M(i_1+i_2)=i}}f(M(j_1+j_2)-M(i+k))\nonumber\\
\noalign{\vglue-5mm}
&\qquad\qquad\qquad\qquad\qquad\qquad{}\cdot p_5((i_1,i_2),(j_1,j_2),k)\nonumber\\
&\qquad\bigg/\sum_{i=0}^{4+u}\;\sum_{\substack{0\le i_1\le i_2\le 9:\\ M(i_1+i_2)=i}}p_5((i_1,i_2),(j_1,j_2),k)
\end{align}
and
\begin{align}\label{eu1(j1,j2,k)-B2}
e_{u,1}((j_1,j_2),k)
&:=\sum_{i=0}^{4+u}\;\sum_{\substack{0\le i_1\le i_2\le 9:\\ M(i_1+i_2)=i}}\;\sum_{l=0}^9 f(M(j_1+j_2+l)-M(i+k))\nonumber\\
\noalign{\vglue-5mm}
&\qquad\qquad\qquad\qquad\qquad\qquad\quad{}\cdot p_6((i_1,i_2),(j_1,j_2),k,l)\nonumber\\
&\qquad\bigg/\sum_{i=0}^{4+u}\;\sum_{\substack{0\le i_1\le i_2\le 9:\\ M(i_1+i_2)=i}}\;\sum_{l=0}^9 p_6((i_1,i_2),(j_1,j_2),k,l),
\end{align}
where $u\in\{0,1\}$ denotes Player's pure strategy ($u=0$ if Player always stands on two-card totals of 5, $u=1$ if Player always draws on two-card totals of 5).  Notice that the denominators of Equation~\eqref{eu0(j1,j2,k)-B2} and Equation~\eqref{eu1(j1,j2,k)-B2} are equal;  we denote their common value by $p_u((j_1,j_2),k)$.

We further define, for $u\in\{0,1\}$ and $0\le j_1\le j_2\le9$ with $M(j_1+j_2)\le7$, 
\begin{align}\label{eu0(j1,j2,e)-B2}
e_{u,0}((j_1,j_2),\varnothing)&:=\sum_{i=5+u}^7\;\sum_{\substack{0\le i_1\le i_2\le9:\\ M(i_1+i_2)=i}} f(M(j_1+j_2)-i)p_4((i_1,i_2),(j_1,j_2))\nonumber\\
&\qquad\bigg/\sum_{i=5+u}^7\;\sum_{\substack{0\le i_1\le i_2\le9:\\ M(i_1+i_2)=i}}p_4((i_1,i_2),(j_1,j_2))
\end{align}
and
\begin{align}\label{eu1(j1,j2,e)-B2}
e_{u,1}((j_1,j_2),\varnothing)
&:=\sum_{i=5+u}^7\;\sum_{\substack{0\le i_1\le i_2\le9:\\ M(i_1+i_2)=i}}\;\sum_{l=0}^9 f(M(j_1+j_2+l)-i)\nonumber\\
\noalign{\vglue-5mm}
&\qquad\qquad\qquad\qquad\qquad\qquad\quad{}\cdot p_5((i_1,i_2),(j_1,j_2),l)\nonumber\\
&\qquad\bigg/\sum_{i=5+u}^7\;\sum_{\substack{0\le i_1\le i_2\le9:\\ M(i_1+i_2)=i}}\;\sum_{l=0}^9
p_5((i_1,i_2),(j_1,j_2),l),
\end{align}
where $u\in\{0,1\}$ has the same interpretation as above.  Notice that the denominators of Equation~\eqref{eu0(j1,j2,e)-B2} and Equation~\eqref{eu1(j1,j2,e)-B2} are equal;  we denote their common value by $p_u((j_1,j_2),\varnothing)$.

We turn to Banker's payoffs in the casino game.  Let us define
\begin{equation}\label{f(x)}
f(x):=\text{sgn}(x)-\alpha\,\bm1_{(0,\infty)}(x)=\begin{cases}1-\alpha&\text{if $x>0$},\\0&\text{if $x=0$},\\-1&\text{if $x<0$}.
\end{cases}
\end{equation}

If Banker has two-card hand $(j_1,j_2)$, where $0\le j_1\le j_2\le9$ and $M(j_1+j_2)\le7$, and Player's third-card value is $k\in\{0,1,\ldots,9\}$, then Banker's standing ($v=0$) and drawing ($v=1$) expectations are, with $f$ as in Equation~\eqref{f(x)},
\begin{align}\label{buv(j1,j2,k)-B2}
&\!\!\! b_{u,v}((j_1,j_2),k)\nonumber\\
&:=E[G_v-\alpha\,\bm1_{\{G_v=1\}}\mid X\in\{0,1,\ldots,4+u\},\,(Y_1,Y_2)=(j_1,j_2),\,X_3=k]\nonumber\\
&\;=E[G_v\mid X\in\{0,1,\ldots,4+u\},\,(Y_1,Y_2)=(j_1,j_2),\,X_3=k]\nonumber\\ 
&\qquad{}-\alpha\,P(G_v=1\mid X\in\{0,1,\ldots,4+u\},\,(Y_1,Y_2)=(j_1,j_2),\,X_3=k)\nonumber\\ 
&\;=e_{u,v}((j_1,j_2),k),\quad u,v\in\{0,1\},
\end{align}
where $u$ denotes Player's pure strategy ($u=0$ if Player always stands at two-card totals of 5, $u=1$ if Player always draws at two-card totals of 5).  Here $100\,\alpha$ is the percent commission on Banker wins.  Throughout we assume that $0\le\alpha\le1/10$.

If Banker has two-card hand $(j_1,j_2)$, where $0\le j_1\le j_2\le9$ and $M(j_1+j_2)\le7$, and Player stands, then Banker's standing ($v=0$) and drawing ($v=1$) expectations are, with $f$ as in Equation~\eqref{f(x)},
\begin{align}\label{buv(j1,j2,e)-B2}
&\!\!\! b_{u,v}((j_1,j_2),\varnothing)\nonumber\\
&:=E[G_v-\alpha\,\bm1_{\{G_v=1\}}\mid X\in\{5+u,\ldots,7\},\,(Y_1,Y_2)=(j_1,j_2),\,X_3=\varnothing]\nonumber\\ 
&\;=E[G_v\mid X\in\{5+u,\ldots,7\},\,(Y_1,Y_2)=(j_1,j_2),\,X_3=\varnothing]\nonumber\\ 
&\qquad{}-\alpha\,P(G_v=1\mid X\in\{5+u,\ldots,7\},\,(Y_1,Y_2)=(j_1,j_2),\,X_3=\varnothing)\nonumber\\  
&\;=e_{u,v}((j_1,j_2),\varnothing),\quad u,v\in\{0,1\},
\end{align}
where $u$ denotes Player's pure strategy, as above.

We now define the payoff bimatrix $(\bm A,\bm B)$ to have $(u,T)$ entry $(a_{u,T},b_{u,T})$ for $u\in\{0,1\}$ and $T\subset\{(j_1,j_2): 0\le j_1\le j_2\le9,\,M(j_1+j_2)\le7\}\times\{0,1,\ldots,9,\varnothing\}$, where
\begin{align*}
b_{u,T}&:=p_u(0)b_u(0)+\sum_{\substack{0\le j_1\le j_2\le 9:\\ M(j_1+j_2)\le7}}\;\sum_{\substack{k\in\{0,1,\ldots,9,\varnothing\}:\\ ((j_1,j_2),k)\in T^c}}p_u((j_1,j_2),k)b_{u,0}((j_1,j_2),k)\\
&\qquad\qquad\qquad{}+\sum_{\substack{0\le j_1\le j_2\le 9:\\ M(j_1+j_2)\le7}}\;\sum_{\substack{k\in\{0,1,\ldots,9,\varnothing\}:\\ ((j_1,j_2),k)\in T}}p_u((j_1,j_2),k)b_{u,1}((j_1,j_2),k),
\end{align*}
using Equations~\eqref{buv(j1,j2,k)-B2} and \eqref{buv(j1,j2,e)-B2} and
\begin{align}\label{p_u(0)b_u(0)-B2}
p_u(0)b_u(0)&:=-\alpha\sum_{j=8}^9\;\sum_{i=0}^{j-1}\;\sum_{\substack{0\le i_1\le i_2\le 9:\\ M(i_1+i_2)=i}}\;\sum_{\substack{0\le j_1\le j_2\le9:\\ M(j_1+j_2)=j}}p_4((i_1,i_2),(j_1,j_2))\nonumber\\
&=-\frac{32\,\alpha\,d^2(37120\,d^2-4044\,d+109)}{(52\,d)_4},
\end{align}
and where $a_{u,T}:=-b_{u,T}$ with $\alpha=0$.  Here $(z)_r:=z(z-1)\cdots(z-r+1)$.  Notice that Equation~\eqref{p_u(0)b_u(0)-B2} does not depend on $u$.

We want to find all Nash equilibria of the casino game \textit{baccara chemin de fer} under Model B2, for all positive integers $d$ and for $0\le\alpha\le1/10$.  We  apply Lemma~\ref{lem:Foster}, Foster's algorithm.  Lemma~\ref{lem:reduction} also applies, reducing the game to $2\times2^{20}$, but that is not needed.  We demonstrate the method by treating the case $d=6$ and $0\le\alpha\le1/10$ in detail.  Then we state results for all $d$.

The first step is to derive a preliminary version of Banker's optimal move at each information set for $\alpha=0$ and for $\alpha=1/10$.  At only three of the $44\times11=484$ information sets does Banker's optimal move differ at $\alpha=0$ and $\alpha=1/10$.  Because $b_{u,v}((j_1,j_2),k)$ is linear in $\alpha$, if the optimal move at $((j_1,j_2),k)$ is the same for $\alpha=0$ and $\alpha=1/10$, then it is also the same for $0\le\alpha\le1/10$.  Results are shown in Table~\ref{tab:reduction-B2}.

The sets $\{p(l): l\in T_{0,1}\}$ and $\{p(l): l\in T_{1,0}\}$ of Lemma~\ref{lem:Foster} are the best-response discontinuities.  In the present setting, the sets $T_{0,1}$ and $T_{1,0}$ depend on $\alpha$, call them $T_{0,1}^\alpha$ and $T_{1,0}^\alpha$.  We call $\{p((j_1,j_2),k): ((j_1,j_2),k)\in T_{0,1}^0\cup T_{0,1}^{1/10}\}$, which has 17 elements, and $\{p((j_1,j_2),k): ((j_1,j_2),k)\in T_{1,0}^0\cup T_{1,0}^{1/10}\}$, which has three elements, \textit{best-response-discontinuity curves}.  The 20 such curves can be evaluated as follows:

\begin{align*}
p((0,3),9)&=\frac{471{,}143+1081\,\alpha}{24 (49{,}118-22{,}303\,\alpha)}, & p((7,8),4)&=\frac{22{,}301+223{,}099\,\alpha}{103{,}799}, \\
p((1,2),9)&=\frac{475{,}514-3219\,\alpha}{24 (48{,}930-22{,}303\,\alpha)}, & p((0,6),\varnothing)&=\frac{477{,}191-54{,}732\,\alpha}{12 (49{,}377-26{,}957\,\alpha)}, \\
p((4,9),9)&=\frac{79{,}051-1643\,\alpha}{4 (49{,}117-22{,}302\,\alpha)}, & p((1,5),\varnothing)&=\frac{474{,}840-49{,}249\,\alpha}{24 (24{,}298-13{,}265\,\alpha)}, \\
p((5,8),9)&=\frac{459{,}978+5089\,\alpha}{24 (48{,}334-21{,}947\,\alpha)}, & p((2,4),\varnothing)&=\frac{486{,}444-56{,}617\,\alpha}{108 (5486-2995\,\alpha)}, \\
p((6,7),9)&=\frac{458{,}114+9555\,\alpha}{2 (589{,}498-267{,}671\,\alpha)}, & p((3,3),\varnothing)&=\frac{17 (4903-621\,\alpha)}{144 (691-377\,\alpha)}, \\
p((2,2),1)&=\frac{732{,}517-127{,}942\,\alpha}{24 (31{,}070-13{,}467\,\alpha)}, & p((7,9),\varnothing)&=\frac{239{,}771-25{,}800\,\alpha}{42 (7027-3824\,\alpha)},\\
p((6,8),1)&=\frac{676{,}141-99{,}462\,\alpha}{24 (31{,}442-13{,}465\,\alpha)}, & p((8,8),\varnothing)&=\frac{78{,}837-8300\,\alpha}{112 (875-478\,\alpha)}, \\
p((7,7),1)&=\frac{7(95{,}873-13{,}162\,\alpha)}{24 (31{,}442-13{,}465\,\alpha)}, & p((1,5),6)&=\frac{348{,}662-715{,}139\,\alpha}{24 (13{,}068-8627\,\alpha)},\\
p((0,5),4)&=\frac{2 (13{,}030+112{,}111\,\alpha)}{102{,}143}, & p((2,4),6)&=\frac{116{,}325-239{,}444\,\alpha}{32 (3272-2191\,\alpha)}, \\
p((6,9),4)&=\frac{29{,}485+222{,}912\,\alpha}{103{,}799}, & p((3,3),6)&=\frac{149{,}704-346{,}703\,\alpha}{288 (561-373\,\alpha)}. 
\end{align*}

\begin{table}[H]
\caption{\label{tab:reduction-B2}Banker's optimal move (preliminary version) in the casino game \textit{baccara chemin de fer} under Model B2 with $d=6$ and with $\alpha=0$ and $\alpha=1/10$, indicated by S (stand) or D (draw), except in the 20 cases indicated by S/D (stand if Player always stands at two-card totals of 5, draw if Player always draws at two-card totals of 5) or D/S (draw if Player always stands at two-card totals of 5, stand if Player always draws at two-card totals of 5) in which it depends on Player's pure strategy.\medskip}
\begin{small}
\begin{center}
\begin{tabular}{cccccccccccccc}
\hline
\noalign{\smallskip}
Banker's &&\multicolumn{11}{c}{Player's third-card value ($\varnothing$ if Player stands)}\\
total & & 0 & 1 & 2 & 3 & 4 & 5 & 6 & 7 & 8 & 9 & $\varnothing$ \\
\noalign{\smallskip} \hline
\noalign{\smallskip}
$0,1,2$    && \cellcolor[gray]{0.85}D & \cellcolor[gray]{0.85}D & \cellcolor[gray]{0.85}D & \cellcolor[gray]{0.85}D & \cellcolor[gray]{0.85}D & \cellcolor[gray]{0.85}D & \cellcolor[gray]{0.85}D & \cellcolor[gray]{0.85}D & \cellcolor[gray]{0.85}D & \cellcolor[gray]{0.85}D & \cellcolor[gray]{0.85}D\\
\noalign{\smallskip} \hline
\noalign{\smallskip}
3&&    \cellcolor[gray]{0.85}D & \cellcolor[gray]{0.85}D & \cellcolor[gray]{0.85}D & \cellcolor[gray]{0.85}D & \cellcolor[gray]{0.85}D & \cellcolor[gray]{0.85}D & \cellcolor[gray]{0.85}D & \cellcolor[gray]{0.85}D & S & $*^1$ & \cellcolor[gray]{0.85}D\\
\noalign{\smallskip} \hline
\noalign{\smallskip}
4&&    S &  $*^2$  & \cellcolor[gray]{0.85}D & \cellcolor[gray]{0.85}D & \cellcolor[gray]{0.85}D & \cellcolor[gray]{0.85}D & \cellcolor[gray]{0.85}D & \cellcolor[gray]{0.85}D & S & S & \cellcolor[gray]{0.85}D\\
\noalign{\smallskip} \hline
\noalign{\smallskip}
5&&    S & S & S & S & $*^3$ & \cellcolor[gray]{0.85}D & \cellcolor[gray]{0.85}D & \cellcolor[gray]{0.85}D & S & S & \cellcolor[gray]{0.85}D\\
\noalign{\smallskip} \hline
\noalign{\smallskip}
6&&    S & S & S & S & S & S & $*^5$ & \cellcolor[gray]{0.85}D & S & S & $*^4$\\
\noalign{\smallskip} \hline
\noalign{\smallskip}
7&     &       S & S & S & S & S & S & S & S & S & S & S\\
\noalign{\smallskip}
\hline
\noalign{\medskip}
\end{tabular}

\begin{tabular}{cccc}
\multicolumn{2}{c}{} & $\alpha=0$ & $\alpha=1/10$\\
\noalign{\smallskip} \hline
\noalign{\smallskip}
$^1\,(3,9)$: & $((0,3),9)$, $((1,2),9)$, $((4,9),9)$,  & \multirow{2}{*}{S/D} & \multirow{2}{*}{S/D} \\
& $((5,8),9)$, $((6,7),9)$ &  &  \\
\noalign{\smallskip} \hline
\noalign{\smallskip}
$^2\,(4,1)$: & $((0,4),1)$, $((1,3),1)$, $((5,9),1)$ & S & S \\
& $((2,2),1)$ & S/D & S \\
& $((6,8),1)$, $((7,7),1)$ & S/D & S/D \\
\noalign{\smallskip} \hline
\noalign{\smallskip}
$^3\,(5,4)$: & $((0,5),4)$, $((6,9),4)$, $((7,8),4)$ & S/D & S/D \\
& $((1,4),4)$, $((2,3),4)$ & S & S \\
\noalign{\smallskip} \hline
\noalign{\smallskip}
$^4\,(6,\varnothing)$: & $((0,6),\varnothing)$, $((1,5),\varnothing)$, $((2,4),\varnothing)$, & \multirow{2}{*}{S/D} & \multirow{2}{*}{S/D} \\
& $((3,3),\varnothing)$, $((7,9),\varnothing)$, $((8,8),\varnothing)$ &  &  \\
\noalign{\smallskip}
\hline
\noalign{\smallskip}
$^5\,(6,6)$: & $((0,6),6)$, $((7,9),6)$, $((8,8),6)$ & D & D \\
& $((1,5),6)$, $((2,4),6)$ & D & D/S \\
& $((3,3),6)$ & D/S & D/S \\
\noalign{\smallskip}
\hline
\end{tabular}
\end{center}
\end{small}
\end{table}

\begin{figure}[ht]
\begin{center}
\includegraphics[width=4.75in]{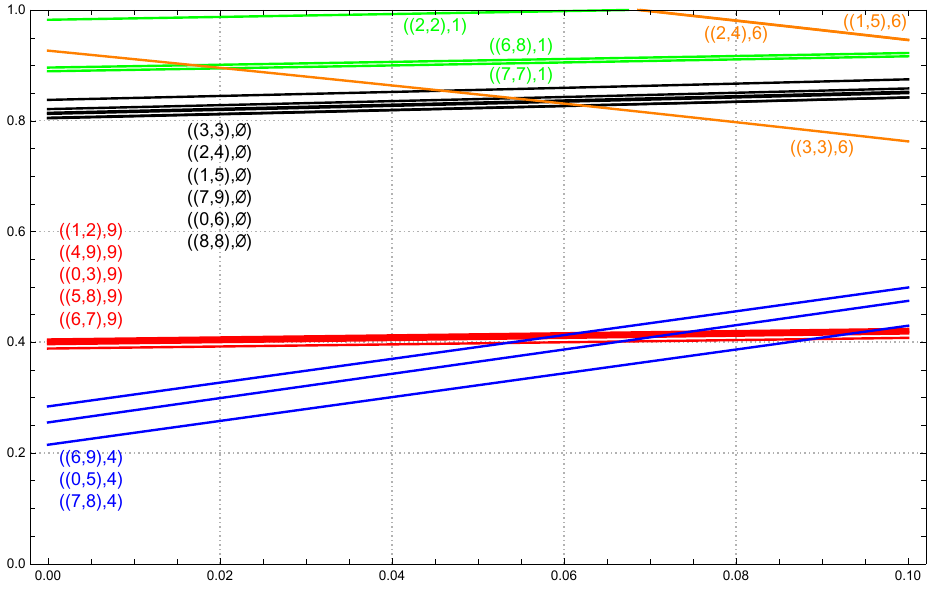}
\vglue5mm
\caption{\label{fig:brdc-B2}The 20 best-response-discontinuity curves for Model B2 and $d=6$  graphed simultaneously as functions of $\alpha\in[0,1/10]$ with range restricted to $[0,1]$.  There are 23 points of intersection.  No two curves of the same color intersect each other.  The labels on the red, blue, and black curves are listed from largest $p$ to smallest $p$.  For example, $p((6,9),4)>p((0,5),4)>p((7,8),4)$.} 
\end{center}
\end{figure}

In Figure~\ref{fig:brdc-B2} these 20 best-response-discontinuity curves are graphed simultaneously.  Notice that $p((0,3),9)$, $p((1,2),9)$, $p((4,9),9)$, $p((5,8),9)$, and $p((6,7),\break 9)$ (red) intersect $p((0,5),4)$, $p((6,9),4)$, and $p((7,8),4)$ (blue); $p((6,8),1)$ and $p((7,7),1)$ (green) and $p((0,6),\varnothing)$, $p((1,5),\varnothing)$, $p((2,4),\varnothing)$, $p((3,3),\varnothing)$, $p((7,9),\break\varnothing)$, and $p((8,8),\varnothing)$ (black) intersect $p((3,3),6)$ (orange).  Thus, there are 23 points of intersection.

We notice also that three of the curves are only partially defined, in that they intersect the horizontal line $p=1$.  These are $p((2,2),1)$ (green) and $p((1,5),6)$ and $p((2,4),6)$ (orange).  They correspond to the three entries in Table~\ref{tab:reduction-B2} that differ at $\alpha=0$ and $\alpha=1/10$.

We are now ready to identify the cases that must be checked for equalizing strategies.  If $r$ is the number of best-response-discontinuity curves and $s$ is the number of points of intersection of these curves, then there are $r+2s$ $\alpha$-intervals and $s$ $\alpha$-values that must be checked for equalizing strategies.  When $d=6$ we have seen that $r=20$ and $s=23$, hence there are 66 $\alpha$-intervals and 23 $\alpha$-values that require attention.  We have summarized these 89 cases in Tables~\ref{tab:66cases} and \ref{tab:23cases}.

Let us provide more detail on Table~\ref{tab:66cases}.  For each best-response-discontinuity curve, if it is intersected by $m$ other best-response-discontinuity curves, that divides the interval $[0,1/10]$ into $m+1$ subintervals, each of which contributes a row to Table~\ref{tab:66cases}.  The Banker strategy for a given row can be deduced from Lemma~\ref{lem:Foster}.  Let us consider row 44.  The Banker strategy DDDDD-SSS-DDD-MSSSSD-DDD, together with Table~\ref{tab:reduction-B2}, allows us to evaluate Player's $2\times2$ payoff matrix, which is
\begin{equation*}
\bm A=\bordermatrix{& \text{B: S at }((0,6),\varnothing)    & \text{B: D at }((0,6),\varnothing) \cr
\text{P: S at 5} & -\frac{22{,}721{,}165{,}499}{1{,}525{,}814{,}595{,}305} & -\frac{3{,}606{,}648{,}223}{305{,}162{,}919{,}061} \cr
\noalign{\smallskip}
\text{P: D at 5} & -\frac{2{,}716{,}895{,}133}{217{,}973{,}513{,}615}   & -\frac{20{,}151{,}297{,}323}{1{,}525{,}814{,}595{,}305} \cr},
\end{equation*}

\noindent
and solving
\begin{equation*}
\begin{pmatrix}1&-1\end{pmatrix}\bm A\begin{pmatrix}1-q\\ q\end{pmatrix}=0
\end{equation*}
yields the equalizing strategy
\begin{equation}\label{q-left}
q=77{,}143{,}741/121{,}269{,}912.
\end{equation}

For row 45, Banker's strategy differs from that in row 44 only at $((3,3),6)$ and we obtain
\begin{equation*}
\bm A=\bordermatrix{& \text{B: S at }((0,6),\varnothing)    & \text{B: D at }((0,6),\varnothing) \cr
\text{P: S at 5} & -\frac{22{,}707{,}392{,}731}{1{,}525{,}814{,}595{,}305} & -\frac{18{,}019{,}468{,}347}{1{,}525{,}814{,}595{,}305} \cr
\noalign{\smallskip}
\text{P: D at 5} & -\frac{19{,}019{,}357{,}419}{1{,}525{,}814{,}595{,}305}   & -\frac{20{,}152{,}388{,}811}{1{,}525{,}814{,}595{,}305} \cr},
\end{equation*}
which yields the equalizing strategy 
\begin{equation}\label{q-right}
q=76{,}834{,}069/121{,}269{,}912.
\end{equation}

Next we provide more detail on Table~\ref{tab:23cases}.  In row 18, corresponding to the intersection of $p((0,6),\varnothing)$ and $p((3,3),6)$, which occurs at
\begin{equation*}
\alpha_0:=\frac{16{,}145{,}999{,}279-\sqrt{226{,}436{,}619{,}657{,}206{,}227{,}489}}{17{,}712{,}223{,}814}\approx0.0620017, 
\end{equation*}
the Banker strategy DDDDD-SSS-DDD-MSSSSD-DDM, together with Table~\ref{tab:reduction-B2}, allow us to evaluate Player's $2\times4$ payoff matrix, which is 
\begin{equation*}
\bordermatrix{& \text{B: SS} & \text{B: SD} & \text{B: DS} & \text{B: DD} \cr
\text{P: S at 5} & -\frac{22{,}707{,}392{,}731}{1{,}525{,}814{,}595{,}305} & -\frac{22{,}721{,}165{,}499}{1{,}525{,}814{,}595{,}305} & -\frac{18{,}019{,}468{,}347}{1{,}525{,}814{,}595{,}305} & -\frac{3{,}606{,}648{,}223}{305{,}162{,}919{,}061} \cr
\noalign{\smallskip}
\text{P: D at 5} & -\frac{19{,}019{,}357{,}419}{1{,}525{,}814{,}595{,}305} & -\frac{2{,}716{,}895{,}133}{217{,}973{,}513{,}615} & -\frac{20{,}152{,}388{,}811}{1{,}525{,}814{,}595{,}305} & -\frac{20{,}151{,}297{,}323}{1{,}525{,}814{,}595{,}305} \cr},
\end{equation*}
where, for example, the Banker strategy SD means S at $((0,6),\varnothing)$ and D at $((3,3),6)$.

There are exactly four equalizing strategies with supports of size two, namely
\begin{align}\label{(1-q,0,q,0)}
&(1-q,0,q,0),\qquad q=76{,}834{,}069/121{,}269{,}912, \\ \label{(1-q,0,0,q)}
&(1-q,0,0,q),\qquad q=76{,}834{,}069/120{,}960{,}240, \\ \label{(0,1-q,q,0)}
&(0,1-q,q,0),\qquad q=77{,}143{,}741/121{,}579{,}584, \\ \label{(0,1-q,0,q)}
&(0,1-q,0,q),\qquad q=77{,}143{,}741/121{,}269{,}912. 
\end{align}

\begin{table}[H]
\caption{\label{tab:66cases}The 66 cases that must be checked for an equalizing strategy, under Model B2 with $d=6$ and $0\le\alpha\le1/10$.  The Banker strategy DDDDD-SSS-DDD-MSSSSD-DDD means draw at $((0,3),9)$, $((1,2),9)$, $((4,9),9)$, $((5,8),9)$, and $((6,7),9)$; stand at $((2,2),1)$, $((6,8),1)$, and $((7,7),1)$;  draw at $((0,5),4)$, $((6,9),4)$, and $((7,8),4)$; mix at $((0,6),\varnothing)$, stand at $((1,5),\varnothing)$, $((2,4),\varnothing)$, $((3,3),\varnothing)$, and $((7,9),\varnothing)$, and draw at $((8,8),\varnothing)$; and draw at $((1,5),6)$, $((2,4),6)$, and $((3,3),6)$.  (For other choices of $((j_1,j_2),k)$ see Table~\ref{tab:reduction-B2}.) Only in cases 44 and 45 is there an equalizing strategy.\medskip}
\catcode`@=\active\def@{\phantom{0}}
\tabcolsep=1.8mm
\begin{center}
\begin{small}
\begin{tabular}{cccc}
\hline
\noalign{\smallskip}
\multirow{2}{*}{case} & Player         & approximate & Banker strategy at \\
     &   $p$          & $\alpha$ interval & $(3,9)$, $(4,1)$, $(5,4)$, $(6,\varnothing)$, $(6,6)$ \\
\noalign{\smallskip}
\hline
\noalign{\smallskip}
@1 & $p((0,3),9)$ & $[0.,0.0589814)$ & MSSDD-SSS-DDD-SSSSSS-DDD \\
@2 & $p((0,3),9)$ & $(0.0589814,0.0720302)$ & MSSDD-SSS-DSD-SSSSSS-DDD \\
@3 & $p((0,3),9)$ & $(0.0720302,0.0943555)$ & MSSDD-SSS-SSD-SSSSSS-DDD \\
@4 & $p((0,3),9)$ & $(0.0943555,0.1]$ & MSSDD-SSS-SSS-SSSSSS-DDD \\
@5 & $p((1,2),9)$ & $[0.,0.0616535)$ & DMDDD-SSS-DDD-SSSSSS-DDD \\
@6 & $p((1,2),9)$ & $(0.0616535,0.0746382)$ & DMDDD-SSS-DSD-SSSSSS-DDD \\
@7 & $p((1,2),9)$ & $(0.0746382,0.0970241)$ & DMDDD-SSS-SSD-SSSSSS-DDD \\
@8 & $p((1,2),9)$ & $(0.0970241,0.1]$ & DMDDD-SSS-SSS-SSSSSS-DDD \\
@9 & $p((4,9),9)$ & $[0.,0.0601033)$ & DSMDD-SSS-DDD-SSSSSS-DDD \\
10 & $p((4,9),9)$ & $(0.0601033,0.0730711)$ & DSMDD-SSS-DSD-SSSSSS-DDD \\
11 & $p((4,9),9)$ & $(0.0730711,0.0953236)$ & DSMDD-SSS-SSD-SSSSSS-DDD \\
12 & $p((4,9),9)$ & $(0.0953236,0.1]$ & DSMDD-SSS-SSS-SSSSSS-DDD \\
13 & $p((5,8),9)$ & $[0.,0.0574359)$ & SSSMD-SSS-DDD-SSSSSS-DDD \\
14 & $p((5,8),9)$ & $(0.0574359,0.0705339)$ & SSSMD-SSS-DSD-SSSSSS-DDD \\
15 & $p((5,8),9)$ & $(0.0705339,0.0928460)$ & SSSMD-SSS-SSD-SSSSSS-DDD \\
16 & $p((5,8),9)$ & $(0.0928460,0.1]$ & SSSMD-SSS-SSS-SSSSSS-DDD \\
17 & $p((6,7),9)$ & $[0.,0.0533616)$ & SSSSM-SSS-DDD-SSSSSS-DDD \\
18 & $p((6,7),9)$ & $(0.0533616,0.0665524)$ & SSSSM-SSS-DSD-SSSSSS-DDD \\
19 & $p((6,7),9)$ & $(0.0665524,0.0887637)$ & SSSSM-SSS-SSD-SSSSSS-DDD \\
20 & $p((6,7),9)$ & $(0.0887637,0.1]$ & SSSSM-SSS-SSS-SSSSSS-DDD \\
21 & $p((2,2),1)$ & $[0.,0.0674106)$ & DDDDD-MDD-DDD-DDDDDD-DDS \\
22 & $p((6,8),1)$ & $[0.,0.0169646)$ & DDDDD-SMD-DDD-DDDDDD-DDD \\
23 & $p((6,8),1)$ & $(0.0169646,0.1]$ & DDDDD-SMD-DDD-DDDDDD-DDS \\
24 & $p((7,7),1)$ & $[0.,0.0205398)$ & DDDDD-SSM-DDD-DDDDDD-DDD \\
25 & $p((7,7),1)$ & $(0.0205398,0.1]$ & DDDDD-SSM-DDD-DDDDDD-DDS \\
26 & $p((0,5),4)$ & $[0.,0.0665524)$ & SSSSS-SSS-MSD-SSSSSS-DDD \\
27 & $p((0,5),4)$ & $(0.0665524,0.0705339)$ & SSSSD-SSS-MSD-SSSSSS-DDD \\
28 & $p((0,5),4)$ & $(0.0705339,0.0720302)$ & SSSDD-SSS-MSD-SSSSSS-DDD \\
29 & $p((0,5),4)$ & $(0.0720302,0.0730711)$ & DSSDD-SSS-MSD-SSSSSS-DDD \\
30 & $p((0,5),4)$ & $(0.0730711,0.0746382)$ & DSDDD-SSS-MSD-SSSSSS-DDD \\
31 & $p((0,5),4)$ & $(0.0746382,0.1]$ & DDDDD-SSS-MSD-SSSSSS-DDD \\
32 & $p((6,9),4)$ & $[0.,0.0533616)$ & SSSSS-SSS-DMD-SSSSSS-DDD \\
33 & $p((6,9),4)$ & $(0.0533616,0.0574359)$ & SSSSD-SSS-DMD-SSSSSS-DDD \\
34 & $p((6,9),4)$ & $(0.0574359,0.0589814)$ & SSSDD-SSS-DMD-SSSSSS-DDD \\
\noalign{\smallskip}
\hline
\end{tabular}
\end{small}
\end{center}
\end{table}

\setcounter{table}{1}

\begin{table}[H]
\caption{(\textit{continued}).\medskip}
\catcode`@=\active\def@{\phantom{D}}
\tabcolsep=1.8mm
\begin{center}
\begin{small}
\begin{tabular}{cccc}
\hline
\noalign{\smallskip}
\multirow{2}{*}{case} & Player         & approximate & Banker strategy at \\
    &   $p$          & $\alpha$ interval & $(3,9)$, $(4,1)$, $(5,4)$, $(6,\varnothing)$, $(6,6)$ \\
\noalign{\smallskip}
\hline
\noalign{\smallskip}
35 & $p((6,9),4)$ & $(0.0589814,0.0601033)$ & DSSDD-SSS-DMD-SSSSSS-DDD \\
36 & $p((6,9),4)$ & $(0.0601033,0.0616535)$ & DSDDD-SSS-DMD-SSSSSS-DDD \\
37 & $p((6,9),4)$ & $(0.0616535,0.1]$ & DDDDD-SSS-DMD-SSSSSS-DDD \\
38 & $p((7,8),4)$ & $[0.,0.0887637)$ & SSSSS-SSS-SSM-SSSSSS-DDD \\
39 & $p((7,8),4)$ & $(0.0887637,0.0928460)$ & SSSSD-SSS-SSM-SSSSSS-DDD \\
40 & $p((7,8),4)$ & $(0.0928460,0.0943555)$ & SSSDD-SSS-SSM-SSSSSS-DDD \\
41 & $p((7,8),4)$ & $(0.0943555,0.0953236)$ & DSSDD-SSS-SSM-SSSSSS-DDD \\
42 & $p((7,8),4)$ & $(0.0953236,0.0970241)$ & DSDDD-SSS-SSM-SSSSSS-DDD \\
43 & $p((7,8),4)$ & $(0.0970241,0.1]$ & DDDDD-SSS-SSM-SSSSSS-DDD \\
\cellcolor[gray]{0.85}44 & \cellcolor[gray]{0.85}$p((0,6),\varnothing)$ & \cellcolor[gray]{0.85}$[0.,0.0620017)$ & \cellcolor[gray]{0.85}DDDDD-SSS-DDD-MSSSSD-DDD \\
\cellcolor[gray]{0.85}45 & \cellcolor[gray]{0.85}$p((0,6),\varnothing)$ & \cellcolor[gray]{0.85}$(0.0620017,0.1]$ & \cellcolor[gray]{0.85}DDDDD-SSS-DDD-MSSSSD-DDS \\
46 & $p((1,5),\varnothing)$ & $[0.,0.0572395)$ & DDDDD-SSS-DDD-DMSSDD-DDD \\
47 & $p((1,5),\varnothing)$ & $(0.0572395,0.1]$ & DDDDD-SSS-DDD-DMSSDD-DDS \\
48 & $p((2,4),\varnothing)$ & $[0.,0.0541199)$ & DDDDD-SSS-DDD-DDMSDD-DDD \\
49 & $p((2,4),\varnothing)$ & $(0.0541199,0.1]$ & DDDDD-SSS-DDD-DDMSDD-DDS \\
50 & $p((3,3),\varnothing)$ & $[0.,0.0458777)$ & DDDDD-SSS-DDD-DDDMDD-DDD \\
51 & $p((3,3),\varnothing)$ & $(0.0458777,0.1]$ & DDDDD-SSS-DDD-DDDMDD-DDS \\
52 & $p((7,9),\varnothing)$ & $[0.,0.0583077)$ & DDDDD-SSS-DDD-DSSSMD-DDD \\
53 & $p((7,9),\varnothing)$ & $(0.0583077,0.1]$ & DDDDD-SSS-DDD-DSSSMD-DDS \\
54 & $p((8,8),\varnothing)$ & $[0.,0.0622043)$ & DDDDD-SSS-DDD-SSSSSM-DDD \\
55 & $p((8,8,\varnothing)$ & $(0.0622043,0.1]$ & DDDDD-SSS-DDD-SSSSSM-DDS \\
56 & $p((1,5),6)$ & $(0.0689443,0.1]$ & DDDDD-SDD-DDD-DDDDDD-MSS \\
57 & $p((2,4),6)$ & $(0.0686285,0.1]$ & DDDDD-SDD-DDD-DDDDDD-DMS \\
58 & $p((3,3),6)$ & $[0.,0.0169646)$ & DDDDD-SDD-DDD-DDDDDD-DDM \\
59 & $p((3,3),6)$ & $(0.0169646,0.0205398)$ & DDDDD-SSD-DDD-DDDDDD-DDM \\
60 & $p((3,3),6)$ & $(0.0205398,0.0458777)$ & DDDDD-SSS-DDD-DDDDDD-DDM \\
61 & $p((3,3),6)$ & $(0.0458777,0.0541199)$ & DDDDD-SSS-DDD-DDDSDD-DDM \\
62 & $p((3,3),6)$ & $(0.0541199,0.0572395)$ & DDDDD-SSS-DDD-DDSSDD-DDM \\
63 & $p((3,3),6)$ & $(0.0572395,0.0583077)$ & DDDDD-SSS-DDD-DSSSDD-DDM \\
64 & $p((3,3),6)$ & $(0.0583077,0.0620017)$ & DDDDD-SSS-DDD-DSSSSD-DDM \\
65 & $p((3,3),6)$ & $(0.0620017,0.0622043)$ & DDDDD-SSS-DDD-SSSSSD-DDM \\
66 & $p((3,3),6)$ & $(0.0622043,0.1]$ & DDDDD-SSS-DDD-SSSSSS-DDM \\
\noalign{\smallskip}
\hline
\end{tabular}
\end{small}
\end{center}
\end{table}

To summarize then, if $\alpha\ne\alpha_0$, then there is a unique Nash equilibrium $(\bm p,\bm q)=((1-p,p),(1-q,q))$, with
\begin{equation}\label{p(0,6,10)(d=6)}
p=p((0,6),\varnothing)=\frac{477{,}191 - 54{,}732\,\alpha}{12 (49{,}377 - 26{,}957\,\alpha)}
\end{equation}
and with $q$ as in Equation~\eqref{q-left} if $0\le\alpha<\alpha_0$ and $q$ as in Equation~\eqref{q-right} if $\alpha_0<\alpha\le1/10$.  If $\alpha=\alpha_0$, uniqueness fails.  Nash equilibria include $(\bm p,\bm q)=((1-p,p),\bm q)$, with $p$ as in Equation~\eqref{p(0,6,10)(d=6)} and $\bm q$ as in Equations~\eqref{(1-q,0,q,0)}--\eqref{(0,1-q,0,q)}.  Moreover, any mixture of these four Nash equilibria is a Nash equilibrium.

\begin{table}[htb]
\caption{\label{tab:23cases}The 23 intersections that must be checked for an equalizing strategy, under Model B2 with $d=6$ and $0\le\alpha\le1/10$.  The meaning of the Banker strategies is as in Table~\ref{tab:66cases}.  Only in case 18 are there equalizing strategies.\medskip}
\catcode`@=\active\def@{\phantom{0}}
\tabcolsep=1mm
\begin{center}
\begin{small}
\begin{tabular}{cccc}
\hline
\noalign{\smallskip}
\multirow{2}{*}{case} & \multirow{2}{*}{intersecting curves} & \multirow{2}{*}{approx.~$\alpha$} & Banker strategy at \\
    &                     &                  & $(3,9)$, $(4,1)$, $(5,4)$, $(6,\varnothing)$, $(6,6)$ \\
\noalign{\smallskip}
\hline
\noalign{\smallskip}
@1 & $p((0, 3), 9)$ \& $p((0, 5), 4)$ & 0.0720302 & MSSDD-SSS-MSD-SSSSSS-DDD \\ 
@2 & $p((0, 3), 9)$ \& $p((6, 9), 4)$ & 0.0589814 & MSSDD-SSS-DMD-SSSSSS-DDD \\ 
@3 & $p((0, 3), 9)$ \& $p((7, 8), 4)$ & 0.0943555 & MSSDD-SSS-SSM-SSSSSS-DDD \\ 
@4 & $p((1, 2), 9)$ \& $p((0, 5), 4)$ & 0.0746382 & DMDDD-SSS-MSD-SSSSSS-DDD \\ 
@5 & $p((1, 2), 9)$ \& $p((6, 9), 4)$ & 0.0616535 & DMDDD-SSS-DMD-SSSSSS-DDD \\ 
@6 & $p((1, 2), 9)$ \& $p((7, 8), 4)$ & 0.0970241 & DMDDD-SSS-SSM-SSSSSS-DDD \\ 
@7 & $p((4, 9), 9)$ \& $p((0, 5), 4)$ & 0.0730711 & DSMDD-SSS-MSD-SSSSSS-DDD \\ 
@8 & $p((4, 9), 9)$ \& $p((6, 9), 4)$ & 0.0601033 & DSMDD-SSS-DMD-SSSSSS-DDD \\ 
@9 & $p((4, 9), 9)$ \& $p((7, 8), 4)$ & 0.0953236 & DSMDD-SSS-SSM-SSSSSS-DDD \\ 
10 & $p((5, 8), 9)$ \& $p((0, 5), 4)$ & 0.0705339 & SSSMD-SSS-MSD-SSSSSS-DDD \\ 
11 & $p((5, 8), 9)$ \& $p((6, 9), 4)$ & 0.0574359 & SSSMD-SSS-DMD-SSSSSS-DDD \\ 
12 & $p((5, 8), 9)$ \& $p((7, 8), 4)$ & 0.0928460 & SSSMD-SSS-SSM-SSSSSS-DDD \\ 
13 & $p((6, 7), 9)$ \& $p((0, 5), 4)$ & 0.0665524 & SSSSM-SSS-MSD-SSSSSS-DDD \\ 
14 & $p((6, 7), 9)$ \& $p((6, 9), 4)$ & 0.0533616 & SSSSM-SSS-DMD-SSSSSS-DDD \\ 
15 & $p((6, 7), 9)$ \& $p((7, 8), 4)$ & 0.0887637 & SSSSM-SSS-SSM-SSSSSS-DDD \\ 
16 & $p((6, 8), 1)$ \& $p((3, 3), 6)$ & 0.0169646 & DDDDD-SMD-DDD-DDDDDD-DDM \\ 
17 & $p((7, 7), 1)$ \& $p((3, 3), 6)$ & 0.0205398 & DDDDD-SSM-DDD-DDDDDD-DDM \\ 
\cellcolor[gray]{0.85}18 & \cellcolor[gray]{0.85}$p((0, 6), \varnothing)$ \& $p((3, 3), 6)$ & \cellcolor[gray]{0.85}0.0620017 & \cellcolor[gray]{0.85}DDDDD-SSS-DDD-MSSSSD-DDM \\ 
19 & $p((1, 5), \varnothing)$ \& $p((3, 3), 6)$ & 0.0572395 & DDDDD-SSS-DDD-DMSSDD-DDM \\ 
20 & $p((2, 4), \varnothing)$ \& $p((3, 3), 6)$ & 0.0541199 & DDDDD-SSS-DDD-DDMSDD-DDM \\ 
21 & $p((3, 3), \varnothing)$ \& $p((3, 3), 6)$ & 0.0458777 & DDDDD-SSS-DDD-DDDMDD-DDM \\ 
22 & $p((7, 9), \varnothing)$ \& $p((3, 3), 6)$ & 0.0583077 & DDDDD-SSS-DDD-DSSSMD-DDM \\ 
23 & $p((8, 8), \varnothing)$ \& $p((3, 3), 6)$ & 0.0622043 & DDDDD-SSS-DDD-SSSSSM-DDM \\
\noalign{\smallskip}
\hline
\end{tabular}
\end{small}
\end{center}
\end{table}

Notice that the Nash equilibrium with $p$ as in Equation~\eqref{p(0,6,10)(d=6)} and $\bm q$ as in Equation~\eqref{(1-q,0,q,0)} coincides with the one from row 45 of Table~\ref{tab:66cases}, the Nash equilibrium with $p$ as in Equation~\eqref{p(0,6,10)(d=6)} and $\bm q$ as in Equation~\eqref{(0,1-q,0,q)} coincides with the one from row 44 of Table~\ref{tab:66cases}.  The two others, with $p$ as in Equation~\eqref{p(0,6,10)(d=6)} and $\bm q$ as in Equation~\eqref{(1-q,0,0,q)} or Equation~\eqref{(0,1-q,q,0)}, are new.

The next step is to verify the three conditions in part (b) of Lemma~\ref{lem:Foster}.  The first condition is easy because the work has already been done in checking for equalizing strategies.  Consider $[0,1/10]\times[0,1]$ minus the union of the 20 best-response-discontinuity curves, as shown in Figure~\ref{fig:brdc-B2}.  It is the union of 43 disjoint connected open regions.  The best response $T^\alpha(p)$ is constant on each of these regions, so we can see that the entries of $\bm A$ corresponding to column $T^\alpha(p)$ have already been computed in analyzing the 66 cases of Table~\ref{tab:66cases}.  

The second condition is easiest because the strategy is the same for $p=0$ and all $\alpha$.  (The case $d=1$ is an exception, and it can be checked separately.)

The third condition is a little more involved because of the three best-response-discontinuity curves that intersect $p=1$.  They divide $[0,1/10]$ into four intervals, and the third condition can be confirmed for each.

This completes the analysis of the case $d=6$.  Statistics for other values of $d$ are shown in Table~\ref{tab:other-d}.

\begin{table}[htb]
\caption{\label{tab:other-d}Dependence on $d$ of various quantities associated with the casino game under Model B2 with $0\le\alpha\le1/10$.  Column (a) contains the number of best-response-discontinuity curves; column (b) contains the number of points of intersection of these curves; column (c) contains the number of these curves that intersect $p=1$ or $p=0$; column (d) contains the number of $\alpha$-intervals that must be checked for equalizing strategies;  and column (e) contains the number of $\alpha$-values at which the Nash equilibrium is nonunique.\medskip}
\catcode`@=\active\def@{\phantom{0}}
\begin{small}
\begin{center}
\begin{tabular}{ccccccccccccc}
\hline
\noalign{\smallskip}
$d$ & (a) & (b) & (c) & (d) & (e) && $d$ & (a) & (b) & (c) & (d) & (e) \\
\noalign{\smallskip}
\hline
\noalign{\smallskip}
@1 & 26 & 13 & @4 & @52 & 2 &@@& 13 & 28 & 28 & 9 & @84 & 0 \\
@2 & 23 & 30 & @3 & @83 & 1 && 14 & 28 & 27 & 9 & @82 & 0 \\
@3 & 22 & @9 & @4 & @40 & 0 && 15--16 & 28 & 30 & 7 & @88 & 0 \\
@4 & 21 & 19 & @4 & @59 & 1 && 17 & 28 & 35 & 7 & @98 & 0 \\
@5 & 21 & 24 & @4 & @69 & 1 && 18 & 28 & 35 & 6 & @98 & 0 \\
@6 & 20 & 23 & @3 & @66 & 1 && 19 & 28 & 36 & 6 & 100 & 0 \\
@7 & 26 & 24 & @9 & @74 & 1 && 20 & 28 & 44 & 6 & 116 & 0 \\
@8 & 28 & 34 & 10 & @96 & 2 && 21--37 & 28 & 46 & 6 & 120 & 0 \\
@9 & 28 & 39 & @9 & 106 & 2 && 38--44 & 28 & 41 & 6 & 110 & 0 \\
10 & 28 & 28 & @8 & @84 & 2 && 45 & 28 & 38 & 6 & 104 & 0 \\
11 & 28 & 31 & @9 & @90 & 2 && 46--76 & 28 & 36 & 6 & 100 & 0 \\
12 & 28 & 26 & @9 & @80 & 0 && $\ge77$ & 28 & 37 & 6 & 102 & 0 \\
\noalign{\smallskip}
\hline
\end{tabular}
\end{center}
\end{small}
\end{table}

Next, we summarize results under Model B2 for all $d\ge1$.  See Table~\ref{tab:summary-B2/B3}.  First, all Nash equilibria $(\bm p,\bm q)=((1-p,p),\bm q)$ have the same $p$, namely
\begin{equation}\label{p-B2}
p=p((0,6),\varnothing)=\frac{(8\,d - 1) (12\,d - 1) (24\,d - 1)-2\,\alpha\,d (128\,d^2-8\,d+1)}{2\,d (1408\,d^2-220\,d+9)-2\,\alpha\,d (768\,d^2-116\,d+5)},
\end{equation}
which generalizes Equation~\eqref{p(0,6,10)(d=6)}.

Table~\ref{tab:uniqueNE-B2} indicates the strategies on which Banker mixes, with drawing probability $q$.
For $d=1$,
\begin{align}\label{1q1}
q&=290{,}383/450{,}072\text{ if }\alpha\in[0,\alpha_1),\\ \label{1q2}
q&=288{,}499/450{,}072\text{ if }\alpha\in(\alpha_1,\alpha_2),\\ \label{1q3}
q&=40{,}811/64{,}296\text{ if }\alpha\in(\alpha_2,1/10].
\end{align}

For $d=2$, 
\begin{align}\label{2q1}
q&=2{,}591{,}845/4{,}119{,}192\text{ if }\alpha\in[0,\alpha_3),\\ \label{2q2}
q&=872{,}479/1{,}373{,}064\text{ if }\alpha\in(\alpha_3,1/10].
\end{align}

For $d=3$, $q$ is as in Equation~\eqref{4567q2} if $\alpha\in[0,1/10]$.  

\begin{table}[H]
\caption{\label{tab:summary-B2/B3}Banker's optimal move in the casino game \textit{baccara chemin de fer} under Model B2 (or B3) for all $d\ge1$ and $0\le\alpha\le1/10$, indicated by S (stand) or D (draw).  For the five asterisks, see Table~\ref{tab:uniqueNE-B2} (or Table~\ref{tab:existNE-B3}).\medskip}
\begin{small}
\begin{center}
\begin{tabular}{cccccccccccccc}
\hline
\noalign{\smallskip}
Banker's &&\multicolumn{11}{c}{Player's third-card value ($\varnothing$ if Player stands)}\\
total & & 0 & 1 & 2 & 3 & 4 & 5 & 6 & 7 & 8 & 9 & $\varnothing$ \\
\noalign{\smallskip} \hline
\noalign{\smallskip}
$0,1,2$    && \cellcolor[gray]{0.85}D & \cellcolor[gray]{0.85}D & \cellcolor[gray]{0.85}D & \cellcolor[gray]{0.85}D & \cellcolor[gray]{0.85}D & \cellcolor[gray]{0.85}D & \cellcolor[gray]{0.85}D & \cellcolor[gray]{0.85}D & \cellcolor[gray]{0.85}D & \cellcolor[gray]{0.85}D & \cellcolor[gray]{0.85}D\\
\noalign{\smallskip} \hline
\noalign{\smallskip}
3&&    \cellcolor[gray]{0.85}D & \cellcolor[gray]{0.85}D & \cellcolor[gray]{0.85}D & \cellcolor[gray]{0.85}D & \cellcolor[gray]{0.85}D & \cellcolor[gray]{0.85}D & \cellcolor[gray]{0.85}D & \cellcolor[gray]{0.85}D & $*$ & \cellcolor[gray]{0.85}D & \cellcolor[gray]{0.85}D\\
\noalign{\smallskip} \hline
\noalign{\smallskip}
4&&    S &  $*$  & \cellcolor[gray]{0.85}D & \cellcolor[gray]{0.85}D & \cellcolor[gray]{0.85}D & \cellcolor[gray]{0.85}D & \cellcolor[gray]{0.85}D & \cellcolor[gray]{0.85}D & S & S & \cellcolor[gray]{0.85}D\\
\noalign{\smallskip} \hline
\noalign{\smallskip}
5&&    S & S & S & S & $*$ & \cellcolor[gray]{0.85}D & \cellcolor[gray]{0.85}D & \cellcolor[gray]{0.85}D & S & S & \cellcolor[gray]{0.85}D\\
\noalign{\smallskip} \hline
\noalign{\smallskip}
6&&    S & S & S & S & S & S & $*$ & \cellcolor[gray]{0.85}D & S & S & $*$\\
\noalign{\smallskip} \hline
\noalign{\smallskip}
7&     &       S & S & S & S & S & S & S & S & S & S & S\\
\noalign{\smallskip}
\hline
\noalign{\medskip}
\end{tabular}
\end{center}
\end{small}
\end{table}

For $4\le d\le7$,
\begin{align}\label{4567q1}
q&=\frac{368{,}640\,d^4-68{,}624\,d^3-2168\,d^2+981\,d-48}{8\,d (52\,d-5) (1408\,d^2-220\,d+9)}\text{ if }\alpha\in[0,\alpha_0),\\ \label{4567q2}
q&=\frac{367{,}104\,d^4-68{,}000\,d^3-2228\,d^2+981\,d-48}{8\,d (52\,d-5) (1408\,d^2-220\,d+9)}\text{ if }\alpha\in(\alpha_0,1/10].
\end{align}

For $d=8,9$, 
\begin{equation}\label{89q1}
q=\frac{367{,}616\,d^4-67{,}728\,d^3-2416\,d^2+1015\,d-51}{8\,d (52\,d-5) (1408\,d^2-220\,d+9)}\text{ if }\alpha\in[0,\alpha_4),\\
\end{equation}
$q$ is as in Equation~\eqref{4567q1} if $\alpha\in(\alpha_4,\alpha_0)$, and $q$ is as in Equation~\eqref{4567q2} if $\alpha\in(\alpha_0,1/10]$.

For $d=10,11$, 
\begin{equation}\label{1011q1}
q=\frac{366{,}592\,d^4-67{,}344\,d^3-2456\,d^2+1017\,d-51}{8\,d (52\,d-5) (1408\,d^2-220\,d+9)}\text{ if }\alpha\in[0,\alpha_5),\\ \end{equation}
$q$ is as in Equation~\eqref{89q1} if $\alpha\in(\alpha_5,\alpha_0)$, and
\begin{equation}\label{1011q3}
q=\frac{366{,}080\,d^4-67{,}104\,d^3-2476\,d^2+1015\,d-51}{8\,d (52\,d-5) (1408\,d^2-220\,d+9)}\text{ if }\alpha\in(\alpha_0,1/10].
\end{equation}

Finally, for $d\ge12$, $q$ is as in Equation~\eqref{1011q1} if $\alpha\in[0,1/10]$.  

We can obtain the uniqueness of the Nash equilibrium for each $d=1,2,\ldots,76$.  For $d\ge77$, we observe that the best-response-discontinuity curves are ordered in a way that does not depend on $d$.  The six curves corresponding to $(4,1)$ intersect the six partial curves corresponding to $(6,6)$, and $p((1,5),6)$ intersects $p((2,4),6)$.  Thus, there are 37 points of intersection for all $d\ge77$.  With this information we can apply Foster's algorithm with a variable $d$ to get the desired uniqueness.

At each of the exceptional points $\alpha_0$ ($4\le d\le11$), $\alpha_1$ and $\alpha_2$ ($d=1$), $\alpha_3$ ($d=2$), $\alpha_4$ ($d=8,9$), and $\alpha_5$ ($d=10,11$), there are exactly four Nash equilibria with Banker equilibrium strategy having support size 2, just as we saw in the case $d=6$.  We leave the evaluation of the various mixing probabilities to the reader.

We have established the following theorem.

\begin{theorem}\label{thm:uniqueNE-B2}
Consider the casino game \textit{baccara chemin de fer} under Model B2 with $d$ a positive integer and $0\le\alpha\le1/10$.  With rare exceptions, there is a unique Nash equilibrium.  Player's equilibrium strategy is to draw at $5$ with probability as in Equation~\eqref{p-B2}.  Banker's equilibrium strategy is as in Tables~\ref{tab:summary-B2/B3} and \ref{tab:uniqueNE-B2}.  The number of exceptions is two if $d\in\{1,8,9,10,11\}$, one if $d\in\{2,4,5,6,7\}$, and none otherwise.  For each of these exceptional values of $\alpha$, there are four Banker equilibrium strategies of support size $2$.
\end{theorem}

Let us briefly compare the Nash equilibrium of the casino game (Theorem~\ref{thm:uniqueNE-B2}) with that of the parlor game (Ethier and G\'amez, 2013), under Model B2 in both cases.  We also compare them in the limit as $d\to\infty$.

In the casino game, Player's mixing probability (i.e., Player's probability of drawing at two-card totals of 5) is as in Equation~\eqref{p-B2}, which depends explicitly on $d$ and $\alpha$.  Banker's mixing probability (i.e., Banker's probability of drawing at $((0,6),\varnothing)$) depends on $d$ and is a step function in $\alpha$ with zero, one, or two discontinuities (zero, hence no $\alpha$ dependence, if $d=3$ or $d\ge12$).  In the limit as $d\to\infty$, Player's mixing probability converges to 

\begin{equation}\label{p-A1}
p=\frac{9-\alpha}{11-6\,\alpha}, 
\end{equation}
while Banker's mixing probability converges to $179/286$.  It follows that Banker's limiting probability of drawing at $(6,\varnothing)$, including $((0,6),\varnothing)$ and $((8,8),\varnothing)$, is
\begin{equation}\label{q-A1}
q=\frac12\,\frac{179}{286}+\frac{1}{16}=\frac{859}{2288},
\end{equation}
and we recognize Equations~\eqref{p-A1} and \eqref{q-A1} as the parameters of the Model A1 Nash equilibrium.

In the parlor game, the results of the preceding paragraph apply with $\alpha=0$.

\begin{table}[H]
\caption{\label{tab:uniqueNE-B2}$\alpha$-intervals where the Nash equilibrium under Model B2 is unique. $\alpha_0$ (resp., $\alpha_1$, $\alpha_2$, $\alpha_3$, $\alpha_4$, $\alpha_5$) is the $\alpha\in(0,1/10)$ at which $p((0,6),\varnothing)$ intersects $p((3,3),6)$ (resp., $p((5,8),8)$, $p((6,7),8)$, $p((0,5),4)$, $p((1,4),4)$, $p((2,3),4)$).  Also, $\alpha_0\approx0.0203752$, $0.0455422$, $0.0620017$, $0.0736066$, $0.0822287$, $0.0888871$, $0.0941842$, $0.0984987$ ($d=4,5,\ldots,11$), $\alpha_1\approx0.0286666$, 
$\alpha_2\approx0.0353207$, $\alpha_3\approx0.0243989$, $\alpha_4\approx0.0203533$, $0.0740412$ ($d=8,9$), and 
$\alpha_5\approx0.0492165$, $0.0889241$ ($d=10,11$).  See Table~\ref{tab:summary-B2/B3} for the full Banker strategies.\medskip}
\catcode`@=\active\def@{\phantom{0}}
\catcode`#=\active\def#{\phantom{$^0$}}
\begin{center}
\begin{small}
\begin{tabular}{cccc}
\hline
\noalign{\smallskip}
\multirow{2}{*}{$d$} & $\alpha$ & Banker strategy at & mixing \\
    & interval & $(3,8)$, $(4,1)$, $(5,4)$, $(6,\varnothing)$, $(6,6)$ & probability \\
\noalign{\smallskip} \hline
\noalign{\smallskip}
\multirow{3}{*}{1} & $[0,\alpha_1)$ & SSDSS-SSSSDD-SSSSS-MSSSSD-DDDSDD & \eqref{1q1} \\
  & $(\alpha_1,\alpha_2)$ & SSDDS-SSSSDD-SSSSS-MSSSSD-DDDSDD & \eqref{1q2} \\
  & $(\alpha_2,1/10]$ & SSDDD-SSSSDD-SSSSS-MSSSSD-DDDSDD & \eqref{1q3} \\  
\noalign{\smallskip} \hline
\noalign{\smallskip}
\multirow{2}{*}{2} & $[0,\alpha_3)$ & SSSSS-SSSSDD-DSSSD-MSSSSD-DDDSDD & \eqref{2q1} \\
  & $(\alpha_3,1/10]$ & SSSSS-SSSSDD-SSSSD-MSSSSD-DDDSDD & \eqref{2q2} \\
\noalign{\smallskip} \hline
\noalign{\smallskip}
3 & $[0,1/10]$ & SSSSS-SSSSSS-DSSDD-MSSSSD-DDDSDD & \eqref{4567q2} \\
\noalign{\smallskip} \hline
\noalign{\smallskip}
\multirow{2}{*}{4--7} & $[0,\alpha_0)$ & SSSSS-SSSSSS-DSSDD-MSSSSD-DDDDDD & \eqref{4567q1} \\
     & $(\alpha_0,1/10]$ & SSSSS-SSSSSS-DSSDD-MSSSSD-DDDSDD & \eqref{4567q2} \\
\noalign{\smallskip} \hline
\noalign{\smallskip}
\multirow{3}{*}{$8,9$} & $[0,\alpha_4)$ & SSSSS-SSSSSS-DDSDD-MSSSSD-DDDDDD & \eqref{89q1} \\
      & $(\alpha_4,\alpha_0)$ & SSSSS-SSSSSS-DSSDD-MSSSSD-DDDDDD & \eqref{4567q1} \\
      & $(\alpha_0,1/10]$ & SSSSS-SSSSSS-DSSDD-MSSSSD-DDDSDD & \eqref{4567q2} \\
\noalign{\smallskip} \hline
\noalign{\smallskip}
\multirow{3}{*}{$10,11$} & $[0,\alpha_5)$ & SSSSS-SSSSSS-DDDDD-MSSSSD-DDDDDD & \eqref{1011q1} \\
        & $(\alpha_5,\alpha_0)$ & SSSSS-SSSSSS-DDSDD-MSSSSD-DDDDDD & \eqref{89q1} \\
        & $(\alpha_0,1/10]$ & SSSSS-SSSSSS-DDSDD-MSSSSD-DDDSDD & \eqref{1011q3} \\
\noalign{\smallskip} \hline
\noalign{\smallskip}
$\ge12$ & $[0,1/10]$ & SSSSS-SSSSSS-DDDDD-MSSSSD-DDDDDD & \eqref{1011q1} \\
\noalign{\smallskip}
\hline
\end{tabular}
\end{small}
\end{center}
\end{table}

\section{Model B3}\label{sec:B3}

In this section we study Model B3.  Here cards are dealt without replacement from a $d$-deck shoe, and each of Player and Banker sees the composition of his own two-card hand.  Player has a stand-or-draw decision in the five situations corresponding to a two-card total of 5, and Banker has a stand-or-draw decision in $44\times11=484$ situations (44 compositions corresponding to Banker totals of 0--7, and 11 Player third-card values, 0--9 and $\varnothing$), so \textit{baccara chemin de fer} is a $2^5\times 2^{484}$ bimatrix game.

The $2^5$ pure strategies of Player can be labeled by the numbers 0--31 in binary form.  For example, strategy $19=(10011)_2$ denotes the Player pure strategy of drawing at $(0,5)$, standing at $(1,4)$, standing at $(2,3)$, drawing at $(6,9)$, and drawing at $(7,8)$.  More generally, for each $u\in\{0,1,\ldots,31\}$, write $u=16u_1+8u_2+4u_3+2u_4+u_5=(u_1u_2u_3u_4u_5)_2$, where $u_1,u_2,u_3,u_4,u_5\in\{0,1\}$, and define $S_u$ to be the set of two-card hands at which Player, using pure strategy $u$, draws:
\begin{align*}
S_u&:=\{(i_1,i_2): 0\le i_1\le i_2\le 9,\, M(i_1+i_2)\le4\text{ or }(i_1,i_2)=(0,5)\text{ if }u_1=1\\
&\qquad\qquad\qquad{}\text{or }(i_1,i_2)=(1,4)\text{ if }u_2=1\text{ or }(i_1,i_2)=(2,3)\text{ if }u_3=1\\
&\qquad\qquad\qquad{}\text{or }(i_1,i_2)=(6,9)\text{ if }u_4=1\text{ or }(i_1,i_2)=(7,8)\text{ if }u_5=1\}.
\end{align*}
The complement of $S_u$ with respect to $\{(i_1,i_2): 0\le i_1\le i_2\le 9,\, M(i_1+i_2)\le7\}$, written $S_u^c$, is the set of two-card hands at which Player, using pure strategy $u$, stands.

The random variables $(X_1,X_2)$, $(Y_1,Y_2)$, $X_3$, $Y_3$, $G_0$, and $G_1$, as well as the function $M$, have the same meanings as in Section~\ref{sec:B2}.

We continue to use Equations~\eqref{p4}--\eqref{p6}.

Given a function $f$ on the set of integers, let us define, by analogy with Equations~\eqref{eu0(j1,j2,k)-B2} and \eqref{eu1(j1,j2,k)-B2}, for $u\in\{0,1,\ldots,31\}$, $0\le j_1\le j_2\le9$ with $M(j_1+j_2)\le7$, and $k\in\{0,1,\ldots,9\}$,
\begin{align}\label{eu0(j1,j2,k)-B3}
e_{u,0}((j_1,j_2),k)&:=\sum_{\substack{0\le i_1\le i_2\le 9:\\ (i_1,i_2)\in S_u}}f(M(j_1+j_2)-M(i_1+i_2+k))\nonumber\\
\noalign{\vglue-5mm}
&\qquad\qquad\qquad\qquad\qquad\qquad\quad{}\cdot p_5((i_1,i_2),(j_1,j_2),k)\nonumber\\
&\qquad\quad\bigg/\sum_{\substack{0\le i_1\le i_2\le 9:\\ (i_1,i_2)\in S_u}}p_5((i_1,i_2),(j_1,j_2),k)
\end{align}
and
\begin{align}\label{eu1(j1,j2,k)-B3}
e_{u,1}((j_1,j_2),k)
&:=\sum_{\substack{0\le i_1\le i_2\le 9:\\ (i_1,i_2)\in S_u}}\;\sum_{l=0}^9 f(M(j_1+j_2+l)-M(i_1+i_2+k))\nonumber\\
\noalign{\vglue-5mm}
&\qquad\qquad\qquad\qquad\qquad\qquad\quad{}\cdot p_6((i_1,i_2),(j_1,j_2),k,l)\nonumber\\
&\qquad\quad\bigg/\sum_{\substack{0\le i_1\le i_2\le 9:\\ (i_1,i_2)\in S_u}}\;\sum_{l=0}^9 p_6((i_1,i_2),(j_1,j_2),k,l).
\end{align}
Notice that the denominators of Equation~\eqref{eu0(j1,j2,k)-B3} and Equation~\eqref{eu1(j1,j2,k)-B3} are equal;  we denote their common value by $p_u((j_1,j_2),k)$.

We further define, for $u\in\{0,1,\ldots,31\}$ and $0\le j_1\le j_2\le9$ with $M(j_1+j_2)\le7$, 
\begin{align}\label{eu0(j1,j2,e)-B3}
e_{u,0}((j_1,j_2),\varnothing)&:=\sum_{\substack{0\le i_1\le i_2\le9:\\ (i_1,i_2)\in S_u^c}}f(M(j_1+j_2)-M(i_1+i_2))p_4((i_1,i_2),(j_1,j_2))\nonumber\\
&\qquad\quad\bigg/\sum_{\substack{0\le i_1\le i_2\le9:\\ (i_1,i_2)\in S_u^c}}p_4((i_1,i_2),(j_1,j_2))
\end{align}
and
\begin{align}\label{eu1(j1,j2,e)-B3}
e_{u,1}((j_1,j_2),\varnothing)&:=\sum_{\substack{0\le i_1\le i_2\le9:\\ (i_1,i_2)\in S_u^c}}\;\sum_{l=0}^9 f(M(j_1+j_2+l)-M(i_1+i_2))\nonumber\\
\noalign{\vglue-5mm}
&\qquad\qquad\qquad\qquad\qquad\qquad\quad{}\cdot p_5((i_1,i_2),(j_1,j_2),l)\nonumber\\
&\qquad\quad\bigg/\sum_{\substack{0\le i_1\le i_2\le9:\\ (i_1,i_2)\in S_u^c}}\;\sum_{l=0}^9 p_5((i_1,i_2),(j_1,j_2),l).
\end{align}
Notice that the denominators of Equation~\eqref{eu0(j1,j2,e)-B3} and Equation~\eqref{eu1(j1,j2,e)-B3} are equal;  we denote their common value by $p_u((j_1,j_2),\varnothing)$.
In Equations~\eqref{eu0(j1,j2,k)-B3}--\eqref{eu1(j1,j2,e)-B3}, $u$ denotes Player's pure strategy.

If Banker has two-card hand $(j_1,j_2)$, where $0\le j_1\le j_2\le9$ and $M(j_1+j_2)\le7$, and Player's third-card value is $k\in\{0,1,2,\ldots,9\}$, then Banker's standing ($v=0$) and drawing ($v=1$) expectations are, with $f$ as in Equation~\eqref{f(x)}, 
\begin{align}\label{buv(j1,j2,k)-B3}
&\!\!\! b_{u,v}((j_1,j_2),k)\nonumber\\
&:=E[G_v-\alpha\,\bm1_{\{G_v=1\}}\mid (X_1,X_2)\in S_u,\,(Y_1,Y_2)=(j_1,j_2),\,X_3=k]\nonumber\\ 
&\;=E[G_v\mid (X_1,X_2)\in S_u,\,(Y_1,Y_2)=(j_1,j_2),\,X_3=k]\nonumber\\ 
&\qquad{}-\alpha\,P(G_v=1\mid (X_1,X_2)\in S_u,\,(Y_1,Y_2)=(j_1,j_2),\,X_3=k)\nonumber\\ 
&\;=e_{u,v}((j_1,j_2),k),\qquad u\in\{0,1,\ldots,31\},\,v\in\{0,1\}.
\end{align}
Here $100\,\alpha$ is the percent commission on Banker wins.  Throughout we assume that $0\le\alpha\le1/10$.

If Banker has two-card hand $(j_1,j_2)$, where $0\le j_1\le j_2\le9$ and $M(j_1+j_2)\le7$, and Player stands, then Banker's standing ($v=0$) and drawing ($v=1$) expectations are, with $f$ as in Equation~\eqref{f(x)},
\begin{align}\label{buv(j1,j2,e)-B3}
&\!\!\! b_{u,v}((j_1,j_2),\varnothing)\nonumber\\
&:=E[G_v-\alpha\,\bm1_{\{G_v=1\}}\mid (X_1,X_2)\in S_u^c,\,(Y_1,Y_2)=(j_1,j_2),\,X_3=\varnothing]\nonumber\\ 
&\;=E[G_v\mid (X_1,X_2)\in S_u^c,\,(Y_1,Y_2)=(j_1,j_2),\,X_3=\varnothing]\nonumber\\  
&\qquad{}-\alpha\,P(G_v=1\mid (X_1,X_2)\in S_u^c,\,(Y_1,Y_2)=(j_1,j_2),\,X_3=\varnothing)\nonumber\\  
&\;=e_{u,v}((j_1,j_2),\varnothing),\qquad u\in\{0,1,\ldots,31\},\,v\in\{0,1\}.
\end{align}
In Equations~\eqref{buv(j1,j2,k)-B3} and \eqref{buv(j1,j2,e)-B3}, $u$ denotes Player's pure strategy.  

We now define the payoff bimatrix $(\bm A,\bm B)$ to have $(u,T)$ entry $(a_{u,T},b_{u,T})$ for $u\in\{0,1,\ldots,31\}$ and $T\subset\{(j_1,j_2): 0\le j_1\le j_2\le9,\, M(j_1+j_2)\le7\}\times\{0,1,\ldots,9,\varnothing\}$, where
\begin{align*}
b_{u,T}&:=-\frac{32\,\alpha\,d^2(37{,}120\,d^2-4044\,d+109)}{(52\,d)_4}\\
&\qquad{}+\sum_{\substack{0\le j_1\le j_2\le 9:\\ M(j_1+j_2)\le7}}\;\sum_{\substack{k\in\{0,1,\ldots,9,\varnothing\}:\\ ((j_1,j_2),k)\in T^c}}p_u((j_1,j_2),k)b_{u,0}((j_1,j_2),k)\\
&\qquad{}+\sum_{\substack{0\le j_1\le j_2\le 9:\\ M(j_1+j_2)\le7}}\;\sum_{\substack{k\in\{0,1,\ldots,9,\varnothing\}:\\ ((j_1,j_2),k)\in T}}p_u((j_1,j_2),k)b_{u,1}((j_1,j_2),k),
\end{align*}
using Equations~\eqref{buv(j1,j2,k)-B3} and \eqref{buv(j1,j2,e)-B3} and recalling Equation~\eqref{p_u(0)b_u(0)-B2}, and where $a_{u,T}:=-b_{u,T}$ with $\alpha=0$.

We want to find a Nash equilibrium of the casino game \textit{baccara chemin de fer} under Model B3, for all positive integers $d$ and for $0\le\alpha\le1/10$.  We demonstrate the method by treating the case $d=6$ and $0\le\alpha\le1/10$ in detail.  Then we state results for all $d$.

Let $d=6$.  We begin by applying Lemma~\ref{lem:reduction}, both with $\alpha=0$ and $\alpha=1/10$, reducing the game to $2^5\times2^{20}$, where 20 refers to the same 20 information sets we identified in Model B2.  In fact, if we attempt to derive the analogue of Table~\ref{tab:reduction-B2} under Model B3, we find that it is identical to Table~\ref{tab:reduction-B2}.  But here an S entry, for example, means that S is optimal versus each of Player's $2^5$ pure strategies.  An S/D entry, for example, means that S is optimal versus Player's pure strategy SSSSS ($u=0$) and D is optimal versus Player's pure strategy DDDDD ($u=31$).

We recall that, under Model B2, the support of Banker's unique equilibrium strategy comprises the two pure strategies
\begin{equation}\label{pure-0,1}
\begin{split}
&\text{DDDDD-SSS-DDD-SSSSSD-DDD}\\
&\text{DDDDD-SSS-DDD-DSSSSD-DDD}
\end{split}
\end{equation}
at $((0,3),9)$, $((1,2),9)$, $((4,9),9)$, $((5,8),9)$, $((6,7),9)$; at $((2,2),1)$, $((6,8),1)$, $((7,7),1)$; at $((0,5),4)$, $((6,9),4)$, $((7,8),4)$; at $((0,6),\varnothing)$, $((1,5),\varnothing)$, $((2,4),\varnothing)$, $((3,3),\varnothing)$, $((7,9),\varnothing)$, $((8,8),\varnothing)$; and at $((1,5),6)$, $((2,4),6)$, $((3,3),6)$.  (This assumes $0\le\alpha<\alpha_0$.  For $\alpha_0<\alpha\le1/10$, there is one change: D at $((3,3),6)$ is changed to S.)  

The key idea is quite simple.  We consider the $2^5\times2$ bimatrix game obtained from Model B3 by restricting Banker's pure strategies to these two alternatives.  Reversing the roles of Player and Banker, we then have a $2\times2^5$ bimatrix game to which Foster's algorithm (Lemma~\ref{lem:Foster}) applies.  The resulting Nash equilibrium yields a candidate for a Nash equilibrium under Model B3, which we can then, we hope, confirm.  (The method fails for $d=1$, which must be treated separately.  It works otherwise, but an additional $\alpha$-interval appears when $d=2$, 9, or 12.)

To apply Lemma~\ref{lem:Foster}, we will have to redefine our notation.  Temporarily, Banker is player I and Player is player II.  Let $V_0,V_1\subset\{(j_1,j_2): 0\le j_1\le j_2\le9,\, M(j_1+j_2)\le7\}\times\{0,1,\ldots,9,\varnothing\}$ correspond to the two pure strategies in Display \eqref{pure-0,1}; specifically, $V_0$ and $V_1$ are the collections of information sets at which Banker draws.  Given a function $f$ on the set of integers, let us define, for $u\in\{0,1\}$ and $0\le i_1\le i_2\le9$ with $M(i_1+i_2)\le7$, \newpage
\begin{align}\label{eu0(i1,i2)}
&\!\!\! e_{u,0}(i_1,i_2)\nonumber\\
&:=\bigg[\sum_{\substack{0\le j_1\le j_2\le 9:\\ M(j_1+j_2)\le7,\\ ((j_1,j_2),\varnothing)\in V_u^c}} f(M(j_1+j_2)-M(i_1+i_2))p_4((i_1,i_2),(j_1,j_2))\nonumber\\
&\qquad{}+\sum_{\substack{0\le j_1\le j_2\le 9:\\ M(j_1+j_2)\le7,\\ ((j_1,j_2),\varnothing)\in V_u}}\sum_{l=0}^9 f(M(j_1+j_2+l)-M(i_1+i_2))p_5((i_1,i_2),(j_1,j_2),l)\bigg]\nonumber\\
&\qquad\quad\bigg/\sum_{\substack{0\le j_1\le j_2\le 9:\\ M(j_1+j_2)\le7}} p_4((i_1,i_2),(j_1,j_2))
\end{align}
and 
\begin{align}\label{eu1(i1,i2)}
&\!\!\! e_{u,1}(i_1,i_2)\nonumber\\
&:=\bigg[\sum_{\substack{0\le j_1\le j_2\le 9:\\ M(j_1+j_2)\le7}}\;\sum_{\substack{0\le k\le9:\\ ((j_1,j_2),k)\in V_u^c}} f(M(j_1+j_2)-M(i_1+i_2+k))\nonumber\\
\noalign{\vglue-3mm}
&\qquad\qquad\qquad\qquad\qquad\qquad\qquad\qquad\qquad\qquad{}\cdot p_5((i_1,i_2),(j_1,j_2),k)\nonumber\\
&\qquad{}+\sum_{\substack{0\le j_1\le j_2\le 9:\\ M(j_1+j_2)\le7}}\;\sum_{\substack{0\le k\le9:\\ ((j_1,j_2),k)\in V_u}}\sum_{l=0}^9 f(M(j_1+j_2+l)-M(i_1+i_2+k))\nonumber\\
\noalign{\vglue-3mm}
&\qquad\qquad\qquad\qquad\qquad\qquad\qquad\qquad\qquad\qquad{}\cdot p_6((i_1,i_2),(j_1,j_2),k,l)\bigg]\nonumber\\
&\qquad\quad\bigg/\sum_{\substack{0\le j_1\le j_2\le 9:\\ M(j_1+j_2)\le7}}\;\sum_{k=0}^9 p_5((i_1,i_2),(j_1,j_2),k).
\end{align}
Notice that the denominators of Equation~\eqref{eu0(i1,i2)} and Equation~\eqref{eu1(i1,i2)} are equal; we denote the common value by $p_u(i_1,i_2)$, which does not actually depend on $u$.

If Player has two-card hand $(i_1,i_2)$, where $0\le i_1\le i_2\le9$ and $M(i_1+i_2)\le7$,  Banker's expectation when Player stands ($v=0$) or draws ($v=1$) is, with $f$ as in Equation~\eqref{f(x)},
\begin{equation*}
a_{u,v}^*(i_1,i_2):=e_{u,v}(i_1,i_2),\quad u,v\in\{0,1\},
\end{equation*}
where $u\in\{0,1\}$ denotes Banker's pure strategy.  For convenience, this definition ignores the rule that Player has a choice only with a two-card total of 5.

We can now define the $2\times 2^5$ payoff bimatrix $(\bm A^*,\bm B^*)$ to have $(u,T)$ entry $(a_{u,T}^*,b_{u,T}^*)$ for $u\in\{0,1\}$ and $T\subset\{(0,5),(1,4),(2,3),(6,9),(7,8)\}$, where \newpage
\begin{align*}
a_{u,T}^*&=-\frac{32\,\alpha\,d^2(37{,}120\,d^2-4044\,d+109)}{(52\,d)_4}\nonumber\\
&\qquad{}+\sum_{\substack{0\le i_1\le i_2\le9:\\ M(i_1+i_2)=6,7}}p_u(i_1,i_2)a_{u,0}^*(i_1,i_2)+\sum_{\substack{0\le i_1\le i_2\le9:\\0\le M(i_1+i_2)\le4}}p_u(i_1,i_2)a_{u,1}^*(i_1,i_2)\nonumber\\
&\qquad{}+\sum_{\substack{0\le i_1\le i_2\le9:\\ M(i_1+i_2)=5,\\ (i_1,i_2)\in T^c}}p_u(i_1,i_2)a_{u,0}^*(i_1,i_2)+\sum_{\substack{0\le i_1\le i_2\le9:\\ M(i_1+i_2)=5,\\(i_1,i_2)\in T}}p_u(i_1,i_2)a_{u,1}^*(i_1,i_2),
\end{align*}
and where $b_{u,T}^*:=-a_{u,T}^*$ with $\alpha=0$.  Note that $\bm B^*$ is additive, so this game fits into the framework of Lemma~\ref{lem:Foster}.

We find that $T_{1,0}=\{(0,5),(1,4),(2,3),(6,9),(7,8)\}$ and the best-response discontinuities, which do not depend on $\alpha$, satisfy
\begin{equation*}
0<p^*(2,3)<p^*(1,4)<p^*(0,5)<p^*(7,8)<p^*(6,9)<1.
\end{equation*}
This leads to a Player equilibrium mixed strategy of DMSDD at $(0,5)$, $(1,4)$, $(2,3)$, $(6,9)$, and $(7,8)$, drawing at $(1,4)$ with equalizing probability
\begin{equation}\label{q^*}
q^*=\frac{35{,}003+186{,}672\,\alpha}{576 (130-71\,\alpha)}.
\end{equation}
Banker's equilibrium mixed strategy is DDDDD-SSS-DDD-MSSSSD-DDD at $(3,9)$, $(4,1)$, $(5,4)$, $(6,\varnothing)$, and $(6,6)$, together with Table~\ref{tab:reduction-B2}, where Banker draws at $((0,6),\varnothing)$ with probability 
\begin{equation}\label{p(1,4)}
p^*=p^*(1,4)=18{,}885{,}571/36{,}781{,}056.
\end{equation}

Recall that this was derived from Banker's equilibrium mixed strategy under Model B2 when $0\le\alpha<\alpha_0$.  For $\alpha_0<\alpha\le1/10$, we obtain the same $q^*$ as in Equation~\eqref{q^*}, but now Banker's equilibrium mixed strategy is DDDDD-SSS-DDD-MSSSSD-DDS at $(3,9)$, $(4,1)$, $(5,4)$, $(6,\varnothing)$, and $(6,6)$, together with Table~\ref{tab:reduction-B2}, where Banker draws at $((0,6),\varnothing)$ with probability 
\begin{equation}\label{p(1,4)-2}
p^*=p^*(1,4)=18{,}792{,}835/36{,}781{,}056.
\end{equation}

In both cases we have found Nash equilibria of the $2^5\times2$ bimatrix game obtained by restricting Banker to two specific pure strategies, those that arise from Model B2.  We now return to regarding Player as player I and Banker as player II, so we redefine
\begin{equation}\label{p-B3,d=6}
p=\frac{35{,}003+186{,}672\,\alpha}{576 (130-71\,\alpha)}
\end{equation}
and $q$ as in Equations~\eqref{p(1,4)} and \eqref{p(1,4)-2}, resulting in two versions of $(\bm p,\bm q)$ that we hope to confirm
as Nash equilibria for the $2^5\times2^{20}$ bimatrix game of Model B3 under suitable conditions on $\alpha$.

Let $(\bm A,\bm B)$ be the $2^5\times2^{20}$ payoff bimatrix.  Let $\bm p$ be the Player mixed strategy with $1-p$ and $p$ (as in Equation~\eqref{p-B3,d=6}) at entries $19=(10011)_2$ and $27=(11011)$ of 0--31 (0s elsewhere).  Let $\bm q$ be the Banker mixed strategy with $1-q$ and $q$ (as in Equation~\eqref{p(1,4)}) at entries $1{,}019{,}407=(11111000111000001111)_2$ and $1{,}019{,}663=(11111000111100001111)_2$ of 0--1{,}048{,}575 (0s elsewhere).  Then $(\bm p,\bm q)$ is a Nash equilibrium of $(\bm A,\bm B)$ if and only if
\begin{align}\label{Aq condition}
&\text{entries 19 and 27 (of 0--31) of }\bm A\bm q^\T\text{ are equal and maximal},\\ \label{pB condition}
&\text{entries 1{,}019{,}407 and 1{,}019{,}663 (of 0--1{,}048{,}575) of }\bm p\bm B\nonumber\\
&\qquad\qquad\qquad\qquad\qquad\qquad\qquad\quad\; \text{ are equal and maximal}
\end{align}
Now Condition~\eqref{Aq condition} is automatic by virtue of how $(\bm p,\bm q)$ was chosen, so it remains to verify Condition~\eqref{pB condition}, which concerns only rows 19 and 27 (of 0--31) of $\bm B$.  Let $\bm B^\circ$ be the $2\times2^{20}$ submatrix of $\bm B$ comprising rows 19 and 27, so that Condition~\eqref{pB condition} is equivalent to
\begin{align*}
&\text{entries 1{,}019{,}407 and 1{,}019{,}663 (of 0--1{,}048{,}575) of }(1-p,p)\bm B^\circ\nonumber\\
&\qquad\qquad\qquad\qquad\qquad\qquad\qquad\quad\; \text{are equal and maximal}.
\end{align*}

\begin{figure}[htb]
\begin{center}
\includegraphics[width=4.75in]{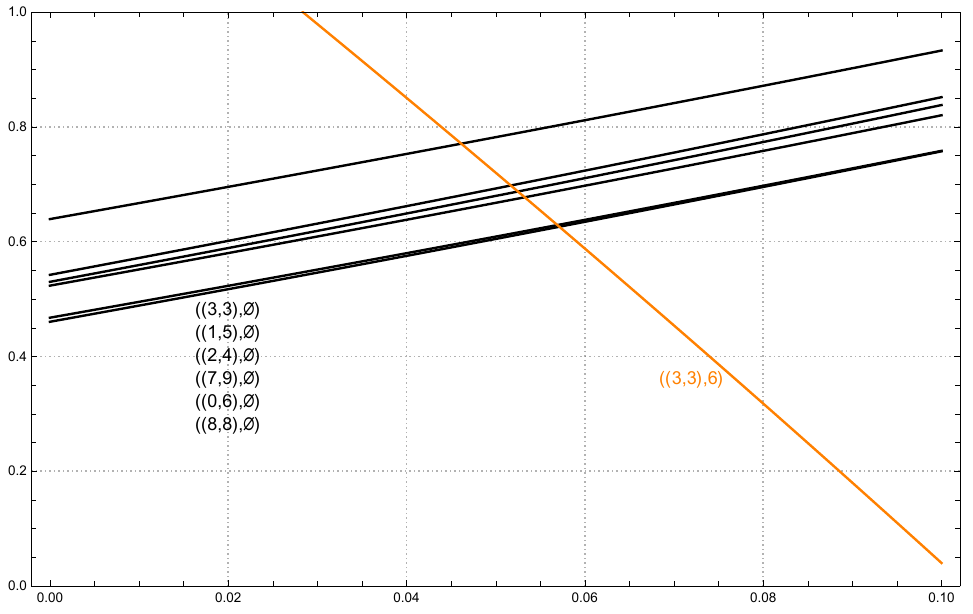}
\vglue4mm
\caption{\label{fig:brdc-B3}The seven best-response-discontinuity curves for Model B3 (restricted to rows 19 and 27 of 0--31) and $d=6$  graphed simultaneously as functions of $\alpha\in[0,1/10]$ with range restricted to $[0,1]$.  There are six points of intersection.} 
\end{center}
\end{figure}

We apply Lemma~\ref{lem:Foster} once again, this time to $(\bm A^\circ,\bm B^\circ)$, where $\bm A^\circ$ is the $2\times2^{20}$ submatrix of $\bm A$ comprising rows 19 and 27 (of 0--31).  We find that $T_{0,1}^\alpha=\{((0,6),\varnothing),((1,5),\varnothing),((2,4),\break\varnothing),((3,3),\varnothing),((7,9),\varnothing),((8,8),\varnothing)\}$, for both $\alpha=0$ and $\alpha=1/10$, while $T_{1,0}^0=\varnothing$ and $T_{1,0}^{1/10}=\{((3,3),6)\}$.  See Figure~\ref{fig:brdc-B3} for the best-response-discontinuity curves.  Furthermore, $p((0,6),\varnothing)$ and $p((3,3),6)$ intersect when $\alpha$ is
\begin{equation*}
\beta_0:=\frac{84{,}325{,}687-\sqrt{6{,}246{,}646{,}053{,}635{,}809}}{92{,}945{,}476}\approx0.0569147.
\end{equation*}
This is enough to conclude that there is a Nash equilibrium for $0\le\alpha<\beta_0$ and another for $\beta_0<\alpha\le1/10$.
Both have the same $\bm p$, namely a $(1-p,p)$ mixture of DSSDD and DDSDD (rows 19 and 27 of 0--31), with $p$ as in Equation~\eqref{p-B3,d=6}.  But the mixture $\bm q$ differs in the two cases.  The first is a $(1-q,q)$ mixture of DDDDD-SSS-DDD-SSSSSD-DDD and DDDDD-SSS-DDD-DSSSSD-DDD (columns 1{,}019{,}407 and 1{,}019{,}663 of 0--1{,}048{,}575), together with Table~\ref{tab:reduction-B2}, with $q$ as in Equation~\eqref{p(1,4)}.  The second is a $(1-q,q)$ mixture of DDDDD-SSS-DDD-SSSSSD-DDS and DDDDD-SSS-DDD-DSSSSD-DDS (columns 1{,}019{,}406 and 1{,}019{,}662 of 0--1{,}048{,}575), together with Table~\ref{tab:reduction-B2}, with $q$ as in Equation~\eqref{p(1,4)-2}.

Finally, just as in Model B2, we obtain multiple Nash equilibria when $\alpha=\beta_0$.  Player strategy DMSDD and Banker strategy DDDDD-SSS-DDD-MSSSSD-DDM, together with Table~\ref{tab:reduction-B2}, allow us to evaluate Player's $2\times4$ payoff matrix, which is 
\begin{small}
\begin{equation*}
\bordermatrix{& \text{B: SS} & \text{B: SD} & \text{B: DS} & \text{B: DD} \cr
\text{P: S at $(1,4)$} & -\frac{3{,}953{,}411{,}487}{305{,}162{,}919{,}061} & -\frac{19{,}769{,}569{,}403}{1{,}525{,}814{,}595{,}305} & -\frac{19{,}423{,}187{,}963}{1{,}525{,}814{,}595{,}305} & -\frac{1{,}765{,}972{,}721}{138{,}710{,}417{,}755} \cr
\noalign{\smallskip}
\text{P: D at $(1,4)$} & -\frac{3{,}878{,}240{,}147}{305{,}162{,}919{,}061} & -\frac{1{,}939{,}185{,}798}{1{,}525{,}814{,}595{,}305} & -\frac{19{,}782{,}952{,}383}{1{,}525{,}814{,}595{,}305} & -\frac{19{,}783{,}609{,}631}{1{,}525{,}814{,}595{,}305} \cr},
\end{equation*} 
\end{small}
where, for example, the Banker strategy SD means S at $((0,6),\varnothing)$ and D at $((3,3),6)$.

Here there are exactly four equalizing strategies with supports of size two, namely
\begin{align*}
&(1-q,0,q,0),\qquad q=18{,}792{,}835/36{,}781{,}056, \\ 
&(1-q,0,0,q),\qquad q=3{,}758{,}567/7{,}337{,}664, \\ 
&(0,1-q,q,0),\qquad q=18{,}885{,}571/36{,}873{,}792, \\ 
&(0,1-q,0,q),\qquad q=18{,}885{,}571/36{,}781{,}056. 
\end{align*}
This completes the description of the Nash equilibria under Model B3 when $d=6$.

Next, we summarize results under Model B3 for all $d\ge1$.  First, there is a Nash equilibrium  $(\bm p,\bm q)$ with Player strategy DMSDD at $(0,5)$, $(1,4)$, $(2,3)$, $(6,9)$, and $(7,8)$, drawing at $(1,4)$ with probability 
\begin{equation}\label{p-B3}
p=p((0,6),\varnothing)=\frac{(12\,d-1)(16\,d^2-14\,d+1)+8\,\alpha\,d (112\,d^2-24\,d+1)}{32\,d^2 (11\,d-1)-16\,d^2\,\alpha (12\,d-1)}
\end{equation}
if $d\ge2$, and 
\begin{equation}\label{p-B3,d=1}
p=p((8,8),\varnothing)=\frac{4 + 203\,\alpha}{2 (38 - 21\,\alpha)}
\end{equation}
if $d=1$.

Table~\ref{tab:existNE-B3} indicates the strategies on which Banker mixes, with drawing probability $q$.  For $d=1$,
\begin{align}\label{1q1-B3}
q&=4519/10{,}716\text{ if }\alpha\in[0,\beta_1),\\ \label{1q2-B3}
q&=3991/10{,}716\text{ if }\alpha\in(\beta_1,1/10].
\end{align}

For $d=2$,
\begin{align}\label{2q1-B3}
q&=17{,}431/64{,}512\text{ if }\alpha\in[0,\beta_2),\\ \label{2q2-B3}
q&=192{,}637/709{,}632\text{ if }\alpha\in(\beta_2,\beta_3),\\ \label{2q3-B3}
q&=65{,}407/236{,}544\text{ if }\alpha\in(\beta_3,1/10].
\end{align}

For $d=3$, $q$ is as in Equation~\eqref{4567q2-B3} if $\alpha\in[0,1/10]$.

For $4\le d\le7$,
\begin{align}\label{4567q1-B3}
q&=\frac{92{,}160\,d^4-120{,}128\,d^3+26{,}336\,d^2-2000\,d+47}{256\,d^2 (11\,d-1) (52\,d-5)}\text{ if }\alpha\in[0,\beta_0),\\ \label{4567q2-B3}
q&=\frac{91{,}776\,d^4-119{,}968\,d^3+26{,}320\,d^2-2000\,d+47}{256\,d^2 (11\,d-1) (52\,d-5)}\text{ if }\alpha\in(\beta_0,1/10].
\end{align}

For $d=8$,
\begin{equation}\label{8q1-B3}
q=\frac{91{,}904\,d^4-119{,}680\,d^3+26{,}064\,d^2-1932\,d+41}{256\,d^2 (11\,d-1) (52\,d-5)}\text{ if }\alpha\in[0,\beta_4),
\end{equation}
$q$ is as in Equation~\eqref{4567q1-B3} if $\alpha\in(\beta_4,\beta_0)$, and $q$ is as in Equation~\eqref{4567q2-B3} if $\alpha\in(\beta_0,1/10]$.

For $d=9$,
\begin{equation}\label{9q1-B3}
q=\frac{91{,}648\,d^4-119{,}488\,d^3+26{,}032\,d^2-1932\,d+41}{256\,d^2 (11\,d-1) (52\,d-5)}\text{ if }\alpha\in[0,\beta_5),
\end{equation}
$q$ is as in Equation~\eqref{8q1-B3} if $\alpha\in(\beta_5,\beta_4)$, $q$ is as in Equation~\eqref{4567q1-B3} if $\alpha\in(\beta_4,\beta_0)$, and $q$ is as in Equation~\eqref{4567q2-B3} if $\alpha\in(\beta_0,1/10]$.

For $d=10,11$, $q$ is as in Equation~\eqref{9q1-B3} if $\alpha\in[0,\beta_5)$, $q$ is as in Equation~\eqref{8q1-B3} if $\alpha\in(\beta_5,\beta_0)$, and 
\begin{equation}\label{1011q3-B3}
q=\frac{91{,}520\,d^4-119{,}520\,d^3+26{,}048\,d^2-1932\,d+41}{256\,d^2 (11\,d-1) (52\,d-5)}\text{ if }\alpha\in(\beta_0,1/10].
\end{equation}

For $d=12$, $q$ is as in Equation~\eqref{9q1-B3} if $\alpha\in[0,\beta_0)$ and
\begin{equation}\label{12q2-B3}
q=1{,}689{,}974{,}681/2{,}989{,}264{,}896\text{ if }\alpha\in(\beta_0,1/10].
\end{equation}

Finally, if $d\ge13$, $q$ is as in Equation~\eqref{9q1-B3} if $\alpha\in[0,1/10]$.

At each of the exceptional points $\beta_0$ ($4\le d\le12$), $\beta_1$ ($d=1$), $\beta_2$ and $\beta_3$ ($d=2$), $\beta_4$ ($d=8,9$), and $\beta_5$ ($d=9,10,11$), there are exactly four Nash equilibria with Banker equilibrium strategy having support size 2, just as we saw in the case $d=6$.  We leave the evaluation of the various mixing probabilities to the reader.

We have established the following theorem.

\begin{theorem}\label{thm:existNE-B3}
Consider the casino game \textit{baccara chemin de fer} under Model B3 with $d$ a positive integer and $0\le\alpha\le1/10$.  There is a Nash equilibrium of the following form.  Player's equilibrium strategy is to draw at $(0,5)$, $(6,9)$, and $(7,8)$, stand at $(2,3)$, and mix at $(1,4)$, drawing with probability as in Equation~\eqref{p-B3} if $d\ge2$, and with probability as in Equation~\eqref{p-B3,d=1} if $d=1$.  Banker's equilibrium strategy is as in Tables~\ref{tab:summary-B2/B3} and \ref{tab:existNE-B3}.  For certain values of $\alpha$ (namely, $\beta_0,\beta_1,\ldots,\beta_5$ of Table~\ref{tab:existNE-B3}), multiple Nash equilibria are known to exist.  The number of such $\alpha$-values is three if $d=9$, two if $d\in\{2,8,10,11\}$, one if $d\in\{1,4,5,6,7,12\}$, and none otherwise.  For each of these $\alpha$-values, Player's equilibrium strategy is as above and Banker has four equilibrium strategies of support size $2$.
\end{theorem}

Let us briefly compare the Nash equilibrium of the casino game (Theorem~\ref{thm:existNE-B3}) with that of the parlor game (Ethier and G\'amez, 2013), under Model B3 in both cases.  We also compare them in the limit as $d\to\infty$.

In the casino game, Player's mixing probability (i.e., Player's probability of drawing at $(1,4)$) is as in Equation~\eqref{p-B3}, which depends explicitly on $d$ and $\alpha$.  
Banker's mixing probability (i.e., Banker's probability of drawing at $((0,6),\varnothing)$) depends on $d$ and is a step function in $\alpha$ with zero, one, two, or three discontinuities (zero, hence no $\alpha$ dependence, if $d=3$ or $d\ge13$).  In the limit as $d\to\infty$, Player's mixing probability converges to $2(3+14\,\alpha)/(11-6\,\alpha)$, while Banker's mixing probability converges to $179/286$.  It follows that Player's limiting probability of drawing at two-card totals of 5, including $(0,5), (1,4)$, $(6,9)$, and $(7,8)$, is
\begin{equation}\label{p-A1-B3}
p=\frac12+\frac18\,\frac{2(3+14\,\alpha)}{11-6\,\alpha}+\frac18+\frac18=\frac{9-\alpha}{11-6\,\alpha},
\end{equation}
and Banker's limiting probability of drawing at $(6,\varnothing)$, including $((0,6),\varnothing)$ and $((8,8),\varnothing)$, is
\begin{equation}\label{q-A1-B3}
q=\frac12\,\frac{179}{286}+\frac{1}{16}=\frac{859}{2288},
\end{equation}
and we recognize Equations~\eqref{p-A1-B3} and \eqref{q-A1-B3} as the parameters of the Model A1 Nash equilibrium.
 
In the parlor game, the results of the preceding paragraph apply with $\alpha=0$.

\section{Summary}

\textit{Baccara chemin de fer} is a classical card game to which game theory applies.  Six models have been proposed; they are obtained by combining either Model A (cards are dealt with replacement) or Model B (cards are dealt without replacement from a $d$-deck shoe) with one of Model 1 (Player and Banker see two-card totals), Model 2 (Player sees two-card totals, Banker sees two-card compositions), or Model 3 (Player and Banker see two-card compositions).  It is further assumed that there is a $100\,\alpha$ percent commission on Banker wins, where $0\le\alpha\le1/10$.  The special case $\alpha=0$ was studied by Ethier and G\'amez (2013).

\begin{table}[H]
\caption{\label{tab:existNE-B3}$\alpha$-intervals where a Nash equilibrium under Model B3 exists. $\beta_1$ is the $\alpha\in(0,1/10)$ at which $p((8,8),\varnothing)$ intersects $p((5,8),8)$.  $\beta_0$ (resp., $\beta_2$, $\beta_3$, $\beta_4$, $\beta_5$) is the $\alpha\in(0,1/10)$ at which $p((0,6),\varnothing)$ intersects $p((3,3),6)$ (resp., $p((6,9),4)$, $p((0,5),4)$, $p((1,4),4)$, $p((2,3),4)$).  Also, $\beta_0\approx0.0120709$, $0.0392263$, $0.0569147$, $0.0693518$, $0.0785740$, $0.0856851$, $0.0913356$, $0.0959335$, $0.0997480$ ($d=4,5,\ldots,12$), $\beta_1\approx0.0655294$, $\beta_2\approx0.0046904$, $\beta_3\approx0.0281623$, $\beta_4\approx0.0244435$, $0.0763202$ ($d=8,9$), and $\beta_5\approx0.0056651$, $0.0544215$, $0.0932285$ ($d=9,10,11$).  See Table~\ref{tab:summary-B2/B3} for the full Banker strategies.\medskip}
\catcode`@=\active\def@{\phantom{0}}
\catcode`#=\active\def#{\phantom{$^0$}}
\begin{center}
\begin{small}
\begin{tabular}{cccc}
\hline
\noalign{\smallskip}
\multirow{2}{*}{$d$}  & $\alpha$ & Banker strategy at & mixing \\
    & interval & $(3,8)$, $(4,1)$, $(5,4)$, $(6,\varnothing)$, $(6,6)$ & probability \\
\noalign{\smallskip} \hline
\noalign{\smallskip}
\multirow{2}{*}{1} & $[0,\beta_1)$    & SSDSD-SSSSDD-SSSSS-SSSSSM-DDDSDD & \eqref{1q1-B3} \\
  & $(\beta_1,1/10]$ & SSDDD-SSSSDD-SSSSS-SSSSSM-DDDSDD & \eqref{1q2-B3} \\  
\noalign{\smallskip} \hline
\noalign{\smallskip}
\multirow{3}{*}{2} & $[0,\beta_2)$       & SSSSS-SSSSDD-DSSDD-MSSSSD-DDDSDD & \eqref{2q1-B3} \\
  & $(\beta_2,\beta_3)$ & SSSSS-SSSSDD-DSSSD-MSSSSD-DDDSDD & \eqref{2q2-B3} \\
  & $(\beta_3,1/10]$    & SSSSS-SSSSDD-SSSSD-MSSSSD-DDDSDD & \eqref{2q3-B3} \\
\noalign{\smallskip} \hline
\noalign{\smallskip}
3 & $[0,1/10]$ & SSSSS-SSSSSS-DSSDD-MSSSSD-DDDSDD & \eqref{4567q2-B3} \\
\noalign{\smallskip} \hline
\noalign{\smallskip}
\multirow{2}{*}{4--7} & $[0,\beta_0)$    & SSSSS-SSSSSS-DSSDD-MSSSSD-DDDDDD & \eqref{4567q1-B3} \\
     & $(\beta_0,1/10]$ & SSSSS-SSSSSS-DSSDD-MSSSSD-DDDSDD & \eqref{4567q2-B3} \\
\noalign{\smallskip} \hline
\noalign{\smallskip}
\multirow{3}{*}{$8$} & $[0,\beta_4)$         & SSSSS-SSSSSS-DDSDD-MSSSSD-DDDDDD & \eqref{8q1-B3} \\
      & $(\beta_4,\beta_0)$ & SSSSS-SSSSSS-DSSDD-MSSSSD-DDDDDD & \eqref{4567q1-B3} \\
      & $(\beta_0,1/10]$    & SSSSS-SSSSSS-DSSDD-MSSSSD-DDDSDD & \eqref{4567q2-B3} \\
\noalign{\smallskip} \hline
\noalign{\smallskip}
\multirow{4}{*}{$9$} & $[0,\beta_5)$           & SSSSS-SSSSSS-DDDDD-MSSSSD-DDDDDD & \eqref{9q1-B3} \\
        & $(\beta_5,\beta_4)$ & SSSSS-SSSSSS-DDSDD-MSSSSD-DDDDDD & \eqref{8q1-B3} \\
        & $(\beta_4,\beta_0)$ & SSSSS-SSSSSS-DSSDD-MSSSSD-DDDDDD & \eqref{4567q1-B3} \\
        & $(\beta_0,1/10]$    & SSSSS-SSSSSS-DSSDD-MSSSSD-DDDSDD & \eqref{4567q2-B3} \\
\noalign{\smallskip} \hline
\noalign{\smallskip}
\multirow{3}{*}{$10,11$} & $[0,\beta_5)$       & SSSSS-SSSSSS-DDDDD-MSSSSD-DDDDDD & \eqref{9q1-B3} \\
        & $(\beta_5,\beta_0)$ & SSSSS-SSSSSS-DDSDD-MSSSSD-DDDDDD & \eqref{8q1-B3} \\
        & $(\beta_0,1/10]$    & SSSSS-SSSSSS-DDSDD-MSSSSD-DDDSDD & \eqref{1011q3-B3} \\
\noalign{\smallskip} \hline
\noalign{\smallskip}
\multirow{2}{*}{$12$} & $[0,\beta_0)$       & SSSSS-SSSSSS-DDDDD-MSSSSD-DDDDDD & \eqref{9q1-B3} \\
        & $(\beta_0,1/10]$ & SSSSS-SSSSSS-DDDDD-MSSSSD-DDDSDD & \eqref{12q2-B3} \\
\noalign{\smallskip}
\hline
\noalign{\smallskip}
$\ge13$ & $[0,1/10]$ & SSSSS-SSSSSS-DDDDD-MSSSSD-DDDDDD & \eqref{9q1-B3} \\
\noalign{\smallskip}
\hline
\end{tabular}
\end{small}
\end{center}
\end{table}

We emphasize Models B2 and B3 in this paper.  Foster’s algorithm, extended to additive $2\times2^n$ bimatrix games, can be applied to Model B2 in a straightforward way, and we obtain, with rare exceptions, a unique Nash equilibrium.  In Model B3 we identify a Nash equilibrium but cannot prove uniqueness.  Here we have a $2^5\times2^{484}$ bimatrix game, which can be reduced to $2^5\times2^{n_d}$, where $20\le n_d\le28$.  We guess that Banker's equilibrium strategy has the same support as it has under Model B2.  We find the Nash equilibrium of the resulting $2^5\times2$ bimatrix game using Foster's algorithm, and finally confirm that this leads to a Nash equilibrium of the full game ($2^5\times2^{n_d}$) by applying Foster's algorithm to the appropriate $2\times 2^{n_d}$ bimatrix game.  The method fails only when $d=1$.  We notice that Player's equilibrium strategy has support independent of $d$ and $\alpha$, which simplifies matters.

\begin{newreferences}

\item Borel, \'E. (1924) \textit{\'El\'ements de la th\'eorie des probabilit\'es}.  Librairie Scientifique, Hermann, Paris, pp. 204--224.

\item Downton, F. and Lockwood, C. (1975) Computer studies of baccarat, I: Chemin-de-fer. \textit{J. Roy. Statist. Soc., Ser. A} \textbf{138} (2) 228--238.

\item Ethier, S. N. and G\'amez, C. (2013) A game-theoretic analysis of \textit{baccara chemin de fer}. \textit{Games} \textbf{4} (4) 711--737.

\item Ethier, S. N. and Lee, J. (2013) Casino \textit{baccara chemin de fer} as a bimatrix game.  Unpublished.  \url{https://arxiv.org/abs/1308.1481v1}.

\item Foster, F. G. (1964) A computer technique for game-theoretic problems I: Chem\-in-de-fer analyzed. \textit{Computer J.} \textbf{7} (2) 124--130.

\item Kemeny, J. G. and Snell, J. L. (1957) Game-theoretic solution of baccarat. \textit{American Mathematical Monthly} \textbf{64} (7) 465--469.

\item Villiod, E. (1906) \textit{La machine \`a voler: \'Etude sur les escroqueries commises dans les cercles \& les casinos}.  Imprimerie Centrale de l'Ouest, Paris.

\end{newreferences}

\end{document}